\documentclass[a4paper,11pt]{article}

\usepackage{amsmath, amssymb}
\usepackage{geometry}
\usepackage{indentfirst}
\usepackage{hyperref}
\usepackage{cite}
\usepackage{cases}
\usepackage{booktabs}
\usepackage{adjustbox}
\usepackage{pdflscape}
\usepackage{color}
\usepackage{amsthm}
\usepackage{citewithnote}

\numberwithin{equation}{section}
\newtheorem{theorem}{Theorem}
\newtheorem{lemma}{Lemma}
\newtheorem{definition}[lemma]{Definition}

\begin{document}

\title{\textbf{Generalized eigenstate thermalization beyond leading order in 2D large $c$ CFTs}}

\author{
Ziyuan Wang$^{a,}$\footnote{First and corresponding author: \href{mailto:simon666_physics@tju.edu.cn}{simon666\_physics@tju.edu.cn}. }\ ,
Shaojie Wang$^{b,}$\footnote{Author: \href{sj-wang25@mails.tsinghua.edu.cn}{sj-wang25@mails.tsinghua.edu.cn}. }
\ and
Jiaju Zhang$^{a,c,}$\footnote{Corresponding author: \href{mailto:jiajuzhang@tju.edu.cn}{jiajuzhang@tju.edu.cn}. }
}

\date{}
\maketitle
\vspace{-9mm}

\begin{center}
{
\small{
$^{a}$\it{Department of Physics, School of Science, Tianjin University, Tianjin 300350, China}

\vspace{1.5mm}

$^{b}$\it{Department of Mathematical Sciences, School of Sciences, Tsinghua University, Beijing 100084, China}

\vspace{1mm}

$^{c}$\it{Center for Joint Quantum Studies, School of Science, Tianjin University, Tianjin 300350, China}
\vspace{1.5mm}
}}
\vspace{10mm}
\end{center}

\begin{abstract}

  We develop a Fock-space formulation that reduces expectation values of vacuum-family quasi-primaries in quantum Korteweg--de Vries (qKdV) eigenstates to a free-bosonic problem, and employ it to test the generalized eigenstate thermalization hypothesis (GETH) in 2D conformal field theories with a large central charge. We prove that the first two qKdV charges $Q_1$ and $Q_3$ lift all degeneracies except along at most countably many curves in the $(c,\,\Delta)$ plane, where $c$ is the central charge and $\Delta$ denotes the highest weight of the conformal family, so that the full-qKdV simultaneous eigenstates are unambiguously determined by $Q_1$ and $Q_3$. Based on this, a comprehensive toolkit for testing GETH within the scope of vacuum-family observables, i.e.~equating qKdV eigenstates and thermal states of the generalized Gibbs ensemble in the sense of expectation values of vacuum-family local observables, is provided in this paper. We argue that GETH is satisfied when the infinite tower of qKdV charges is considered. A more restrictive form of the GETH, namely that all expectations of local observables of a definite level should be expressible in terms of the first few qKdV charges, is directly falsified at subleading order in the large-$c$ expansion. Specifically, at levels 6 and 8 in the vacuum family, expectation values of quasi-primaries in simultaneous eigenstates of the qKdV charges are not determined by any finite number of the lowest-index qKdV charges. The question of whether additional microscopic information about the eigenstate is required is left for future work. 

\end{abstract}
\vspace{0.6\baselineskip}
\noindent
\begin{center}
\begin{minipage}{0.88\textwidth}
\end{minipage}
\end{center}
\vspace{0.8\baselineskip}

\baselineskip 18pt
\thispagestyle{empty}

\tableofcontents\clearpage

\section{Introduction}\label{sec:introduction}
The eigenstate thermalization hypothesis (ETH) posits that individual energy eigenstates are locally equivalent to the corresponding thermal ensembles in the thermodynamic limit~\cite{Deutsch:1991msp,Srednicki:1994mfb,Rigol:2007juv,DAlessio:2015qtq}. In some chaotic systems, ETH is found to hold to a certain extent, due to the absence of a sufficient number of conserved quantities~\cite{Deutsch:1991msp,Srednicki:1994mfb}. However, for integrable systems, there are infinitely many conserved quantities, so the Gibbs ensemble should be replaced by the generalized Gibbs ensemble (GGE)~\cite{Rigol:2007hbz}, and ETH is correspondingly generalized to the generalized eigenstate thermalization hypothesis (GETH)~\cite{Cassidy:2011mia}. GETH states that, for a large integrable system, the expectation values of local observables in a simultaneous eigenstate of all the conserved quantities can be described by a GGE characterized by the corresponding conserved quantities~\cite{Cassidy:2011mia}. The validity of GETH depends on the specific integrable system under consideration. Explicit counterexamples were found in some integrable systems with multiple particle species, where local conserved charges do not completely determine local observables~\cite{Pozsgay:2014ex}. On the other hand, a generalized strong ETH was rigorously established for translation-invariant noninteracting integrable systems~\cite{Ishii:2019cwk}.

Traditional ETH was demonstrated in two-dimensional conformal field theory (2D CFT) in the large central charge limit via light-operator two-point functions~\cite{Fitzpatrick:2014vua,Fitzpatrick:2015zha} and entanglement entropy~\cite{Asplund:2014coa,Caputa:2014eta} for highly excited primary states, both agreeing with the thermal results~\cite{DiFrancesco:1997nk,Calabrese:2004eu} at leading order in the large central charge expansion. However, later studies found deviations of highly excited states from the thermal behavior expected by ETH in diagnostics such as R\'enyi entropy, entanglement entropy, relative entropy, and Schatten distance~\cite{Lashkari:2016vgj,Lin:2016dxa,He:2017vyf,Basu:2017kzo,He:2017txy,Guo:2018pvi,Guo:2018fye}; in particular, Ref.~\cite{Guo:2018pvi} found probable violations of the ETH constraints given in Ref.~\cite{Lashkari:2016vgj} for certain highly excited descendant states, highlighting the role of the additional conserved quantum Korteweg--de Vries (qKdV) charges. Recent work also reached similar conclusions~\cite{Besken:2019bsu,Liska:2022vrd,Fukushima:2023svf,Sui:2025qve}. These studies naturally motivate attempts to replace ordinary Gibbs ensembles with GGEs in 2D CFTs~\cite{Mandal:2015jla,Cardy:2015xaa,deBoer:2016bov,He:2017txy} and provide the background for the subsequent development of GETH from ordinary ETH in 2D CFTs~\cite{Dymarsky:2018lhf,Maloney:2018hdg,Maloney:2018yrz,Dymarsky:2018iwx,Brehm:2019fyy,Dymarsky:2019etq,Datta:2019jeo}. In 2D CFTs, which have integrable structures~\cite{Bazhanov:1994ft,Bazhanov:1996dr,Bazhanov:1998dq}, the qKdV charges $\{Q_{2k-1}\}$ serve as generalized thermodynamic variables in the GGEs~\cite{Calabrese:2006rx,Calabrese:2007rg,Maloney:2018yrz}. GETH asserts that, in the thermodynamic limit, any statistically significant simultaneous eigenstate of all qKdV charges can be equivalently described by a corresponding GGE thermal state, in the sense that the two yield identical expectation values for local observables~\cite{Dymarsky:2018lhf,Dymarsky:2019etq}.

The qKdV--GGE--GETH framework underlying this proposal was developed through studies of saddle-point formulation of the qKdV partition function, comparison between the GGE and qKdV eigenstates at large $c$, quasi-modular methods for thermal qKdV correlation functions, sharply peaked qKdV charge distributions, and the large-$c$ generalized partition function with an infinite tower of qKdV chemical potentials~\cite{Dymarsky:2018lhf,Maloney:2018hdg,Maloney:2018yrz,Dymarsky:2018iwx}, culminating in the explicit formulation of GETH and some leading-order expressions for local observable expectation values in terms of the qKdV charges, testing GETH at leading order up to level 10~\cite{Dymarsky:2019etq}. Supporting the typicality assumption underlying GETH~\cite{Dymarsky:2019etq}, typical high-energy states were shown to be thermal-like Virasoro descendants~\cite{Datta:2019jeo}, and the necessity and effectiveness of GETH in 2D CFTs are further argued~\cite{Besken:2019bsu}. Techniques have been subsequently developed for oscillator representations of the Virasoro algebra, zero-mode computations of quasi-primaries and qKdV charges, and qKdV spectra up to the subleading order~\cite{Besken:2019bsu,Dymarsky:2019iny,Dymarsky:2022dhi}.

The validity of GETH at leading order for level 12 and higher is indirectly supported by Refs.~\cite{Novaes:2021vjh,Chen:2024lji}. Closely related developments have studied GGE properties including thermal expectation values of qKdV charges, chemical potentials, modular properties of the partition function, and entanglement entropy~\cite{Dymarsky:2019etq,Dymarsky:2022dhi,Chen:2024lji}. Generalized eigenstate thermalization in 2D CFTs has found applications in AdS$_3$/CFT$_2$, ranging from generalized black-hole ensembles to subsystem quantum-information measures~\cite{Dymarsky:2020tjh,Chen:2024ysb}. In parallel, complementary work has further developed the understanding of GGE properties, particular modular properties of GGEs~\cite{Downing:2021mfw,Downing:2023lnp,Downing:2024nfb,Karimi:2026tvh,Ashok:2026cyi}. 

However, the validity of GETH beyond the leading order of large central charge has been questioned in Ref.~\cite{Novaes:2021vjh}, where a potential violation was found at subleading order in the semiclassical expansion. In this paper, we further investigate the behavior of local-observable expectation values at the subleading order, and thereby examine the applicability of GETH through a fully quantitative and controlled analysis that directly uses the quantum KdV framework and takes into account the infinite tower of qKdV charges.

We discuss GETH in the chiral sector of 2D CFTs; the anti-chiral sector behaves in exactly the same way. In this paper, we restrict the discussion to GETH in the scope of the vacuum family as in previous literature: operators and densities of conserved charges restricted to the vacuum family, and qKdV eigenstates allowed to belong to any non-vacuum conformal families~\cite{Dymarsky:2018lhf,Dymarsky:2019etq,Novaes:2021vjh}. Operators and charges associated with non-vacuum conformal families are beyond the scope of this paper, as no general framework exists for them~\cite{Novaes:2021vjh}, and their characterization implies requirement of parameters beyond the central charge $c$ and highest weight $\Delta$~\cite{Downing:2021mfw}. 

There exist two independent limits: the limit of large central charge, $c\to+\infty$, and the thermodynamic limit $\ell\to+\infty$, where $\ell$ sets the scale of the system. The former limit merely underlies an expansion in $1/c$ and admits a natural interpretation through the AdS$_3$/CFT$_2$ correspondence~\cite{Brown:1986nw,Dymarsky:2019etq}, whereas the latter is understood as a strict limit~\cite{DAlessio:2015qtq,Dymarsky:2019etq,Dymarsky:2022dhi}. The highest weight $\Delta$ and the occupation-number moments $M_{2r-1}=\sum_{k=1}^{\infty} k^{2r-1}n_k$ scale as~\cite{Dymarsky:2019etq,Datta:2019jeo,Dymarsky:2022dhi}
\[
\Delta\sim c\ell^2,\quad M_{2r-1}\sim\ell^{2r}.
\]
where $n_k$ has the same meaning as in Ref.~\cite{Dymarsky:2019etq}, with its physical meaning further illustrated in Section \ref{sec:fock}. The thermodynamic limit indicates that the densities of conserved charges are fixed. Thus, one parametrizes the eigenvalues of the qKdV charges $Q_{2k-1}$ as $\ell^{2k}q_{2k-1}$, with the eigenvalues $q_{2k-1}$ held fixed as $\ell\to+\infty$~\cite{Dymarsky:2019etq}.

More precisely, the necessary and sufficient condition for GETH to hold is expected to be that the expectation values of local observables in simultaneous eigenstates of the qKdV charges in arbitrary highest-weight conformal families be expressible as smooth functions of the qKdV charges, with the plausible assumption that $\{\mu_{2k-1}\}$ can be inversely solved in terms of $\{q_{2k-1}\}$. We first discuss sufficiency, the logic of which can be organized into two steps. In the first step, consider a simultaneous eigenstate $\lvert\alpha\rangle$ of the qKdV charges, with the corresponding eigenvalues $\ell^{2k}q_{2k-1}$, which may also be denoted by a subscript $\alpha$. The task is to find a set of chemical potentials $\{{\mu_{2k-1}}\}$ such that the eigenstate and the corresponding GGE are equivalent. The density matrix of the GGE is
\begin{equation}\label{eqt:rhogge}
\rho_{\mathrm{GGE}}=\frac{1}{Z_{\mathrm{GGE}}}\mathrm{e}^{-\mathop{\sum_{k=1}^{\infty}} \mu_{2k-1}Q_{2k-1}}=\sum_{\alpha'}p_{\alpha'} |\alpha'\rangle \langle\alpha'|,
\end{equation}
with $Z_{\mathrm{GGE}}$ defined as
\begin{equation}\label{eqt:z}
Z_{\mathrm{GGE}}=\mathrm{tr}\left(\mathrm{e}^{-\mathop{\sum_{k=1}^{\infty}} \mu_{2k-1}Q_{2k-1}}\right),
\end{equation}
with $\mu_1=\beta$ and $\ell^2 q_1=E$, where $E$ is the total energy and $\beta$ is the inverse temperature, and with that
\begin{equation}\label{eqt:gcanonical}
p_{\alpha'}=\frac{\mathrm{e}^{-\mathop{\sum_{k=1}^{\infty}} \mu_{2k-1}\ell^{2k}q_{2k-1}(\alpha')}}{\mathrm{tr}\left(\mathrm{e}^{-\mathop{\sum_{k=1}^{\infty}} \mu_{2k-1}Q_{2k-1}}\right)}
\end{equation}
is the probability of the state $|\alpha'\rangle$. And then $q_{2k-1}^{\mathrm{GGE}}=\frac{1}{\ell^{2k}}\,\mathrm{tr}\left(\rho_{\mathrm{GGE}}Q_{2k-1}\right)$, where $q_{2k-1}^{\mathrm{GGE}}$ is used to denote the average of a qKdV-charge density in a thermal state. The ensemble further gives
\begin{equation}\label{eqt:chempot}
q_{2k-1}^{\mathrm{GGE}}=-\frac{1}{\ell^{2k}}\frac{\partial\log Z_{\mathrm{GGE}}}{\partial\mu_{2k-1}},
\end{equation}
which provides the conditions for determining the chemical potentials~\cite{Maloney:2018yrz}. Consider the following truncation: if only the first $r$ chemical potentials are treated as independent unknowns, while those carrying higher indices are zero, then Eq.~(\ref{eqt:chempot}) likewise provides $r$ equations. A further calculation shows that the Jacobian matrix of the map $\{q_{2k-1}\}\mapsto\{\mu_{2k-1}\}$ is given by the covariance matrix of the qKdV charges, which is positive semidefinite; since the thermal state is not an exact energy eigenstate, the determinant is in fact positive, and hence the chemical potentials are uniquely determined. This formulation is extended from the approach given in Ref.~\cite[Page 91--92]{Schekochihin:2026ssm}. This argument is expected to be extended to the case where the complete infinite tower of chemical potentials $\{\mu_{2k-1}\}$ is taken into account, although this remains an open problem in the strict sense~\cite{Downing:2023lnp}.

The second step is in principle to establish the relation between all expectation values of local observables and the qKdV charges under this matching. On the sufficiency side, we assume that the expectation values of local observables $\mathcal{O}$ are expressible as smooth functions $f_{\mathcal{O}}(\{q_{2k-1}\})$ of the qKdV charges in the discussion on the sufficiency side. By an identical transformation, $\langle\mathcal{O}\rangle_{\mathrm{GGE}}=\sum_{\alpha'}p_{\alpha'}\langle\alpha'|\mathcal{O}|\alpha'\rangle=\sum_{\alpha'}p_{\alpha'}f_{\mathcal{O}}(\{q_{2k-1}\}_{\alpha'})$, where $p_{\alpha'}$ is given by Eq.~(\ref{eqt:gcanonical}). Given the elementary result of state concentration in the thermodynamic limit, whereby the mixed states manifest themselves as sharply peaked distributions of qKdV charges~\cite{Maloney:2018hdg,Maloney:2018yrz}, one has $\langle\mathcal{O}\rangle_{\mathrm{GGE}}=f_{\mathcal{O}}(\{q_{2k-1}^{\mathrm{GGE}}\})$, up to omittable thermodynamically subleading orders~\citewithnote{DAlessio:2015qtq}{DAlessio:2015qtq,Dymarsky:2019etq}{Section 4.2}. Setting $\{q_{2k-1}^{\mathrm{GGE}}\}=\{q_{2k-1}\}_\alpha$, one finally obtains
\begin{equation}\label{eqt:geth}
\langle\mathcal{O}\rangle_{\mathrm{GGE}}=\langle\alpha|\mathcal{O}|\alpha\rangle .
\end{equation}

The derivation of necessity is even more straightforward. In fact,
\begin{equation}\label{eqt:ggemu}
\langle\mathcal{O}\rangle_{\mathrm{GGE}}=\mathrm{tr}\left(\mathcal{O}\rho_{\mathrm{GGE}}\right)
\end{equation}
is a smooth function of $\{\mu_{2k-1}\}$ alone, since $\{q_{2k-1}(\alpha')\}$ is known. With Eq.~(\ref{eqt:chempot}), $\langle\mathcal{O}\rangle_{\mathrm{GGE}}$ is a smooth function of the qKdV charges $\{q_{2k-1}\}$ alone, since thermalization requires the conserved quantities to match. Since one needs to obtain Eq.~(\ref{eqt:geth}), there must be
\begin{equation}\label{eqt:gethexp}
\langle\alpha|\mathcal{O}|\alpha\rangle=f_{\mathcal{O}}(\{q_{2k-1}\}),
\end{equation}
up to thermodynamically subextensive terms, where $f_{\mathcal{O}}$ is a smooth mapping.

The above discussion concerns the general form of GETH, named infinite-tower GETH in this paper. Another version of GETH will also be discussed in this paper, which is regarded as a more restrictive form, named truncated GETH (not to be confused with ``strong version of the generalized ETH'' mentioned in Ref.~\cite{Dymarsky:2019etq}). It asserts that the local properties of an eigenstate can be captured by a finite initial segment of the qKdV charges, without requiring the reconstruction of the full microscopic information encoded in the complete qKdV spectrum; in particular, the expectation value of a local quasi-primary operator at a given level should depend only on a finite initial segment of the qKdV charges at the same and smaller levels~\cite{Dymarsky:2019etq}. 

The necessity direction established above is rigorous, whereas the sufficiency side relies on the existence of the inverse map $\{q_{2k-1}\}\mapsto\{\mu_{2k-1}\}$. Since only the quasi-primary components contribute to nonvanishing expectation values in the scope of the vacuum family, it suffices to consider quasi-primary observables, and specifically only those at even levels~\cite{Dymarsky:2019etq}. The subsequent exclusion of the original form of truncated GETH at the $c$-subleading order in this paper is rigorous, and the infinite-tower GETH claim to an arbitrary order in $1/c$ is subject to the assumption of the inverse map.

We next summarize the conventions and notations adopted in this paper. We introduce the notation associated with the classical action-angle variables~\cite{Dymarsky:2022dhi,Maloney:2018yrz}
\begin{equation}\label{eqt:not}
\tilde{c}=\frac{c-1}{24},\quad\tilde{\Delta}=\Delta-\tilde{c},\quad h=\frac{\tilde{\Delta}}{\tilde{c}},\quad M_{p,\,s}=\sum_{k=1}^{\infty}k^p n_k^s,\quad M_p=M_{p,\,1}.
\end{equation}
The quantities $M_{p,\,s}$ are referred to as occupation-number moments, as $m_{p,\,r}$ in Ref.~\cite{Dymarsky:2022dhi}. We use the term ``linear occupation-number moments'' for $M_p$, and ``quadratic occupation-number moments'' for $M_{p,\,2}$ in this article. For the plane-to-cylinder mapping, this paper adopts the existing practice commonly used in the literature on qKdV structure, following BLZ's convention~\cite{Bazhanov:1994ft}, which differs from Ginsparg's~\cite{Ginsparg:1988ui} by a factor of $i$. We also use curly brackets to denote the set of physical quantities corresponding to all indices; the indices in such a set are assumed by default to range over all positive integers.

This paper is organized as follows. In Section \ref{sec:freeboson} we discuss the free-bosonic realization of the Virasoro algebra in detail, as the preparation for the whole formulation. In Section \ref{sec:spectrum} we discuss the general structure of the qKdV charges and its important corollary, namely the first-order corrections to the qKdV eigenstates, as another essential part of the formulation. In Section \ref{sec:expectations} we complete the calculations of the vacuum-family quasi-primary expectations of levels 6 and 8, providing the analytical basis for truncated GETH. In Section \ref{sec:discomp} we analyze the distinguishability by and completeness of qKdV charges, thereby justifying the choice of GGE and providing conceptual background for later argument for the infinite-tower GETH. In Section \ref{sec:geth} we systematically address GETH at different levels of strength and analyze the outcomes for GETH. Various technical details are deferred to the appendices: the Poincar\'e--Birkhoff--Witt (PBW) states are treated in Appendix \ref{app:pbw}, the measure-zero property of degeneracy curves of first two qKdV charges in Appendix \ref{app:nondeg}, distinguishing microstates with qKdV charges in Appendix \ref{app:sol}, and the role of the central charge in GETH in Appendix \ref{app:der}.
\section{Free-bosonic realization of Virasoro algebra}\label{sec:freeboson}
This section provides the preliminaries for the rest of the article. Indeed, the free-bosonic realization provides a foundation for GETH studies~\cite{Dymarsky:2018lhf,Dymarsky:2018iwx,Dymarsky:2019etq,Dymarsky:2022dhi}. More precisely, PBW descendants serve as the leading order of the simultaneous eigenstates of all qKdV charges, as is proved in Appendix \ref{app:pbw}. However, for the subleading orders in $c$ of qKdV charges and quasi-primaries, the representation based on PBW states is not applicable, because PBW states are not precise eigenstates of the leading order of qKdV charges. Ref.~\cite{Dymarsky:2022dhi} circumvented this difficulty by combining the semiclassical quantization of the classical KdV hierarchy with the ODE/IM correspondence, but we adopt a more direct formulation following Refs.~\cite{Besken:2019bsu,Besken:2019jyw} in this paper. As we will see, this approach enables the discussion of quasi-primary expectations and simplifies high-order calculations for a specific qKdV charge, though it might be less suited than ODE/IM correspondence to deriving general qKdV-charge formulae.
\subsection{Fock space}\label{sec:fock}

For a chosen set of complex variables $\{u_k\}$ mentioned in Ref.~\cite{Besken:2019bsu} with $k$ running over all positive integers, one sets
\[
a_k\mapsto-\frac{i}{\sqrt{2}}\frac{\partial}{\partial u_k},\quad a_{-k}\mapsto i\sqrt{2}k u_k,
\]
then
\begin{equation}\label{eqt:com}
[a_m,\,a_n]=m\delta_{m+n,\,0},
\end{equation}
and the formula (2.2) of Ref.~\cite{Besken:2019jyw} becomes
\begin{equation}\label{eqt:rea}
\begin{cases}
\displaystyle L_k=\sqrt{2\tilde{c}}\,(\sqrt{h}+ik)a_k+\frac{1}{2}\sum_{p=1}^{k-1}a_p a_{k-p}+\sum_{p=1}^{\infty}a_{-p}a_{k+p},\\
\displaystyle L_{-k}=\sqrt{2\tilde{c}}\,(\sqrt{h}-ik)a_{-k}+\frac{1}{2}\sum_{p=1}^{k-1}a_{-p} a_{-(k-p)}+\sum_{p=1}^{\infty}a_{-(k+p)}a_{p},\\
\displaystyle L_0=\Delta+\sum_{p=1}^{\infty}a_{-p}a_p.
\end{cases}
\end{equation}

In the large-$c$ limit, $a_k$ approaches $L_k$ up to an overall factor and $O(1/\tilde{c})$ correlations. One can then denote and normalize any Fock state as
\begin{equation}
|\Delta,\,\{n_k\}\rangle_{\mathrm{Fock}}=\prod_{k=1}^{\infty}\frac{(a_{-k})^{n_k}}{\sqrt{k^{n_k}n_k!}}|\Delta\rangle,
\end{equation}
Here, $|\Delta\rangle$ denotes a primary state with highest weight $\Delta$; $n_k$ denotes the number of occurrences of $a_{-k}$ in the Fock state, or equivalently, the number of occurrences of $L_{-k}$ in the corresponding PBW state as in Ref.~\cite{Dymarsky:2019etq}. These Fock states replace the PBW states $|\Delta,\,\{n_k\}\rangle_{\mathrm{PBW}}$ in the calculation below, as the precise simultaneous eigenstates of the leading order of all qKdV charges, which is further explained in Section \ref{sec:note}. For convenience, we may suppress the $\Delta$ label in $|\Delta,\,\{n_k\}\rangle_{\mathrm{Fock}}$ in this paper. One derives
\begin{equation}\label{eqt:creann}
\begin{cases}
a_r|\cdots,\,n_r,\,\cdots\rangle_{\mathrm{Fock}}=\sqrt{rn_r}|\cdots,\,n_r-1,\,\cdots\rangle_{\mathrm{Fock}},\\
a_{-r}|\cdots,\,n_r,\,\cdots\rangle_{\mathrm{Fock}}=\sqrt{r(n_r+1)}|\cdots,\,n_r+1,\,\cdots\rangle_{\mathrm{Fock}},
\end{cases}
\end{equation}
and then
\begin{equation}
a_{-r}a_r|\Delta,\,\{n_k\}\rangle_{\mathrm{Fock}}=rn_r|\Delta,\,\{n_k\}\rangle_{\mathrm{Fock}} .
\end{equation}
It is also worth noting that different Fock states are orthogonal to one another.

\subsection{Scaling of the conformal weight and the occupation-number moments}\label{sec:scalings}
This subsection discusses the scaling of $\tilde{\Delta}$ and $M_{p,\,s}$. As a preliminary remark, we briefly comment on the change of variables $(c,\,\Delta)\to(\tilde{c},\,\tilde{\Delta})$. At leading order in $c$ or $\tilde{c}$, one can directly replace $c$ by $c-1=24\tilde{c}$, since the difference between the two appears only at subleading order. Hence
\begin{equation}\label{eqt:Delta}
\tilde{\Delta}\sim\tilde{c}\ell^2 .
\end{equation}

After introducing the free-bosonic realization, $n_k$ is understood as the occupation number of a bosonic state. According to bosonic statistics,
\begin{equation}\label{eqt:sup}
\langle n_k\rangle\approx\frac{1}{\mathrm{e}^{\frac{\kappa\beta k}{\ell}}-1}\approx\frac{\ell}{\kappa\beta k},
\end{equation}
where 
\begin{equation}
\kappa=1+6t_3 e^*+15t_5(e^*)^2+\cdots
\end{equation}
with $t_{2r-1}=\left(\frac{c\pi^2}{6\beta^2}\right)^{r-1}\frac{\mu^{2r-1}}{\beta}$ as a constant term without $\ell$-scaling given in Ref.~\cite{Dymarsky:2018lhf}. The first approximation holds because the thermodynamic limit makes the chemical-potential terms with higher $\ell$-order in their denominators vanish, while the second is valid for small $k$, or equivalently for $k \langle n_k \rangle$ large enough. Therefore, up to an overall factor,
\[
M_{p,\,s}=\sum_{k=1}^{\infty}k^p n_k^s\sim\ell^s\sum_{k=1}^{\ell}k^{p-s}.
\]
This follows from the fact that the ratio between the integral over $k$ of the summand in the middle expression in Eq.~(\ref{eqt:sup}) and that of the right-hand-side summand cut off at $k=\ell$ approaches a constant at large $\ell$. So, at large $\ell$,
\[
\displaystyle
\sum_{k=1}^{\ell} k^{\alpha}\sim
\begin{cases}
\mathrm{Const},\,&\alpha<-1,\\
\log{\ell},\,&\alpha=-1,\\
\ell^{\alpha+1},\,&\alpha>-1,
\end{cases}
\]
leading to
\begin{equation}\label{eqt:llim}
M_{p,\,s}\sim
\begin{cases}
\ell^s,\,&p<s-1,\\
\ell^s \log{\ell},\,&p=s-1,\\
\ell^{p+1},\,&p>s-1.
\end{cases}
\end{equation}

The scaling with $\tilde{c}$ of $\kappa$ remains unresolved, but it has to be taken to be independent of $\tilde{c}$ in order to ensure the consistency of the $c$-scaling of $M_{2k-1}$~\cite{Dymarsky:2019etq,Dymarsky:2022dhi}. As a supplement, it is worth noting that this scaling is not imposed a priori; rather, the present work adopts it because some physical meaning underlies it~\cite{Dymarsky:2019etq,Datta:2019jeo}. In contrast, a different scaling $M_{2r-1}\sim c\ell^{2r}$ was considered in Ref.~\cite{Guo:2018pvi}.

\section{Spectrum of qKdV charges and their eigenstates}\label{sec:spectrum}

In this section, we reveal general structure of the qKdV charges as a preparation in Sections \ref{sec:genstr} and \ref{sec:note}, and then derive the first-order perturbative corrections to the qKdV eigenstates in Section \ref{sec:cor}. 

From this section on, the notation $|\Delta,\,\{n_k\}\rangle$ and $|\{n_k\}\rangle$ is defined to refer to the qKdV eigenstate corresponding to the zeroth-order Fock state $|\Delta,\,\{n_k\}\rangle_{\mathrm{Fock}}$ or $|\{n_k\}\rangle_{\mathrm{Fock}}$; any $n_k$ or $M_{p,\,s}$ occurring in the perturbative corrections always denotes the occupation numbers and their moments of that Fock state.

\subsection{Spectral structure of qKdV charges}\label{sec:genstr}

qKdV spectra and expectations of quasi-primaries that belong to the vacuum family are expressed in terms of expectation values of products of Virasoro generators~\cite{Bazhanov:1994ft,Dymarsky:2019iny}. According to Eqs.~(\ref{eqt:rea}), these operators of finite descendent level are exact polynomials in $1/\sqrt{\tilde c}$, with the expansion terminating at a finite order. They therefore form a finite-dimensional analytic family in $\epsilon=\tilde{c}^{-\frac{1}{2}}$ near $\epsilon=0$. Hence, the infinitely-perturbative expectation values of quasi-primaries and qKdV charges are exact, without residue terms like $\mathrm{e}^{-\tilde c}$~\cite[Page 121]{Kato:1995pt}.

Furthermore, odd-order corrections correspond to half-integer powers of $\tilde{c}$. According to Eq.~(\ref{eqt:rea}), increasing the perturbative order by one changes the number of oscillators $a_k$ by one, and hence flips the parity of $M_0$ of the state on one side after the operator acts on it. Similarly, odd-order corrections to the quantum states have $M_0$ parity opposite to that of the zeroth-order state, owing to the nonzero-coefficient condition, while even-order corrections have the same parity. Therefore, for odd-order corrections the total perturbative order of the bra, the operator and the ket is odd, and the bra and the operator-acted ket have opposite $M_0$ parity, so that they are orthogonal. Consequently, no term with half-integer powers of $\tilde{c}$ appears in the expansions, so the expectation values of quasi-primaries and qKdV charges are analytic functions of $1/\tilde{c}$, which is their most significant feature.

The leading-order terms of all qKdV charges are correlated. To make the independent aspects of these physical quantities more explicit, one may instead use the variables
\begin{equation}
\delta q_{2k-1}=\frac{q_{2k-1}}{q_1^{k}}-1,
\end{equation}
which are of order $O(\ell^0\tilde{c}^{-1})$, as a measure of the subleading order of the qKdV charges~\cite{Dymarsky:2019etq}. Considering the subleading order~\cite{Dymarsky:2022dhi}, after taking the thermodynamic limit, i.e.~the strict $\ell\to+\infty$ limit, with Eqs.~(\ref{eqt:Delta}) and (\ref{eqt:llim}), they are expressed as
\begin{equation}\label{eqt:dq3}
\delta q_3=\frac{1}{\tilde{c}}\left(\frac{4M_1}{h}+\frac{4M_3}{h^2}\right)+\frac{1}{\tilde{c}^2}\left[-\frac{6M_1^2}{h^2}-\frac{8M_1M_3}{h^3}+\frac{3M_1}{h}+\frac{M_3}{h^2}-\frac{3M_{0,\,2}}{2h}\right]+O(\tilde{c}^{-3}),
\end{equation}
\begin{equation}
\begin{aligned}\label{eqt:dq5}
\delta q_5=&\ \frac{1}{\tilde{c}}\left(\frac{12M_1}{h}+\frac{20M_3}{h^2}+\frac{8M_5}{h^3}\right)+\frac{1}{\tilde{c}^2}\left[-\frac{14M_1^2}{h^2}-\frac{40M_1M_3}{h^3}-\frac{24M_1M_5}{h^4}\right.\\
&\left.+\frac{15M_1}{h}+\frac{55M_3}{3h^2}+\frac{16M_5}{3h^3}-\frac{15M_{0,\,2}}{2h}\right]+O(\tilde{c}^{-3}),
\end{aligned}
\end{equation}
\begin{equation}
\begin{aligned}\label{eqt:dq7}
\delta q_7=&\ \frac{1}{\tilde{c}} \left( \frac{24M_1}{h} + \frac{56M_3}{h^2} + \frac{224M_5}{5h^3} + \frac{64M_7}{5h^4} \right) + \frac{1}{\tilde{c}^2} \left[ -\frac{4M_1^2}{h^2} - \frac{336M_1M_3}{5h^3} - \frac{672M_1M_5}{5h^4}\right.\\
&\left. - \frac{256M_1M_7}{5h^5} + \frac{42M_1}{h} + \frac{112M_3^2}{5h^4} + \frac{266M_3}{3h^2} + \frac{4816M_5}{75h^3} + \frac{1136M_7}{75h^4} - \frac{21M_{0,\,2}}{h} \right]\\
&+O(\tilde{c}^{-3}).
\end{aligned}
\end{equation}
The leading orders on the right-hand sides of Eqs.~(\ref{eqt:dq3})--(\ref{eqt:dq7}) are exactly the same as those in Ref.~\cite{Dymarsky:2018iwx}. Note that the leading order of $\{\delta q_{2k-1}\}\ (k\geq2)$ generally reads, in the thermodynamic limit,~\cite{Dymarsky:2018iwx}
\begin{equation}\label{eqt:m2q}
\delta q_{2k-1}=\frac{1}{\tilde{c}}\left[2k(k-1)\frac{M_1}{h}+\sum_{j=2}^{k}\frac{(2k-1)\sqrt{\pi}\,\Gamma(k+1)}{2\,\Gamma\!\left(j+\frac12\right)\Gamma(k-j+1)}\frac{M_{2j-1}}{h^j}\right]+O(\tilde{c}^{-2}).
\end{equation}

One additionally defines
\begin{equation}
m=\frac{M_1}{\tilde{\Delta}},
\end{equation}
which is also of order $O(\ell^0\tilde{c}^{-1})$. This physical quantity characterizes the degree of descendant character of a Fock state in the thermodynamic limit, and is used as an analytical tool to complete the representation space. However, in a physical system this is not a good quantity, as is elaborated in Section \ref{sec:physical}.

\subsection{Solution to the degeneracy problem for qKdV eigenstates}\label{sec:note}
It is conjectured that the full set of qKdV charges resolves spectral degeneracies in certain minimal models at special values of $(c,\,\Delta)$~\cite{Bazhanov:1994ft}. Whether this conjecture holds in the cases considered in Ref.~\cite{Bazhanov:1994ft}, and whether it extends to generic unitary and non-unitary CFTs, remain open questions. Here, we investigate whether the first two qKdV charges suffice to lift degeneracies for common $(c,\,\Delta)$'s, and conclude the result by the following theorem. 
\begin{theorem}\label{theo:1}
On the plane spanned by $(c,\,\Delta)$, or equivalently $(\tilde{c},\,\tilde{\Delta})$, $Q_1$ and $Q_3$ can lift all the degeneracies except for at most countably infinitely many curves.
\end{theorem}

This subsection outlines the proof strategy, while the complete proof is given in Appendix \ref{app:nondeg}. It is readily seen that the simultaneous eigenstates of $Q_1$ and $Q_3$ are just those of $Q_1$ and $D=\sum_{k=1}^{\infty}L_{-k} L_k$. As is seen in Appendix \ref{app:pbw}, let $\mathbf{D}_N$ denote the matrix of $D$ acting on the subspace of $Q_1$-eigenstates. The first step in solving the eigensystem is to set the determinant
\begin{equation}\label{eqt:*}
\det\left(\mathbf{D}_N-\lambda\mathbf{I}\right)=0,
\end{equation}
and the roots of this equation are exactly the eigenvalues.

The left-hand side of Eq.~(\ref{eqt:*}) for a certain level $N=M_1$ is a polynomial in $\lambda$, so the degenerate condition is that its discriminant equals zero. Therefore, the negation of the statement is to find some $N$ such that the discriminant of $\det\left(\mathbf{D}_N-\lambda\mathbf{I}\right)$ is identically zero, independently of $(c,\,\Delta)$, because there would otherwise be only countably many lines of parameters for degeneracies given by polynomial roots, even when all $N$ from 0 to $+\infty$ are taken into account.

On the other hand, the eigenvalues of a Hermitian operator such as $D$ are real, so the discriminants must be nonnegative. Thus, it would be very strange for the discriminant just to take the boundary at some $N$, in comparison with those at $N=0\sim7$, because that would break the symmetries. Based on this belief, we give the complete proof of Theorem \ref{theo:1} in this paper. 

Specifically, it suffices to find one parameter pair $(c,\,\Delta)$ for each level $N$ for which the joint spectrum of $Q_1$ and $Q_3$ is nondegenerate. Unitarity is not required for the sought point, since the level-$N$ discriminant is a universal polynomial in $(c,\,\Delta)$. We perturb around the point $c=-2$, where it is easy to obtain precise results due to the existence of precise ODE/IM correspondence~\cite{Masoero:2026rvz}. As background, $c=-2$ is also a meeting point for many theoretical models, such as symplectic fermions~\cite{Kausch:2000fu} and uniform spanning trees~\cite{Liu:2024ybm}, as well as desirable properties such as modular invariance of GGEs~\cite{Downing:2024nfb,Karimi:2026tvh}. Applying the perturbation of the qKdV eigenvalue equations around $c=-2$, at which Ref.~\cite{Masoero:2026rvz} exactly diagonalized the first three qKdV charges, we establish, for every $N$, the existence of $(c,\,\Delta)$ for which the joint degeneracies of $Q_1$ and $Q_3$ are lifted. Therefore, for every $N$, the discriminant of $\mathbf{D}_N$'s determinant $\operatorname{disc}_N$ is not the zero polynomial, thereby proving Theorem \ref{theo:1}. Exact degeneracies may nevertheless occur at special values of $(c,\,\Delta)$, whose example is given by Eq.~(\ref{eqt:counter}): precise $c=-2$ regardless of $\Delta$.

For common $(c,\,\Delta)$'s except on the measure-zero set consisting of the degeneracy curves, the eigenstates have distinguishable $(Q_1,\,Q_3)$ eigenvalues, so they are the exact eigenstates of all qKdV charges. In conclusion, a corollary of the theorem is that the expressions of simultaneous eigenstates of $Q_1$ and $Q_3$ given by a fixed formulation with continuous $(c,\,\Delta)$ are exactly those of all qKdV charges.

\subsection{First-order corrections to qKdV eigenstates}\label{sec:cor}

In this subsection, we mainly calculate the first-order perturbative corrections to the simultaneous eigenstates of all qKdV charges based on the result of Section \ref{sec:note}.

In the thermodynamic limit,
\begin{equation}
Q_3=L_0^2+2\sum_{k=1}^{\infty}L_{-k} L_k .
\end{equation}
Noting that the $L_0$ eigenvalue of a Fock state is $\Delta+\sum_{k=1}^\infty kn_k$, just as for a PBW descendant, and that in the same limit this value becomes $\tilde{\Delta}+\sum_{k=1}^\infty kn_k$, one has
\[
\begin{cases}
\displaystyle Q_3=L_0^2+4\tilde{c}\sum_{k=1}^{\infty}(k^2+h)a_{-k}a_k+O(\tilde{c}^0),\\
\displaystyle \ell^4 q_3=\tilde{\Delta}^2+\tilde{c}\left(2h\sum_{k=1}^\infty kn_k+4\sum_{k=1}^{\infty}k(k^2+h)n_k\right)+O(\tilde{c}^0).
\end{cases}
\]

It is, however, more convenient to consider $D$. $D$ and its eigenvalue $d$ are expanded in powers of $\tilde{c}$, as
\begin{numcases}{}
D=\tilde{c}D^{(0)}+\tilde{c}^{\frac{1}{2}}D^{(1)}+\tilde{c}^0 D^{(2)}+\cdots,\\
d=\tilde{c}d^{(0)}+\tilde{c}^{\frac{1}{2}}d^{(1)}+\tilde{c}^0 d^{(2)}+\cdots,
\end{numcases}
with qKdV charges, their eigenvalues and the eigenstates following the same notation. One has
\begin{equation}\label{eqt:l3d}
d^{(0)}=2\left(M_3+hM_1\right),
\end{equation}
and one reads from Eq.~(\ref{eqt:rea}) that
\begin{align}
D^{(1)}=&\ \sqrt{2}\sum_{k=0}^{\infty}\left[\left(\sqrt{h}-ik\right)a_{-k}\left(\frac{1}{2}\sum_{p=1}^{k-1}a_pa_{k-p}+\sum_{p=1}^{\infty}a_{-p}a_{k+p}\right)\right.\nonumber\\
&\left.+\left(\sqrt{h}+ik\right)a_{-k}\left(\frac{1}{2}\sum_{p=1}^{k-1}a_{-p}a_{-(k-p)}+\sum_{p=1}^{\infty}a_{-(k+p)}a_{p}\right)\right]\nonumber\\
=&\ \frac{3}{2}\sqrt{2h}\sum_{p,\,q=1}^{\infty}(a_{-p}a_{-q}a_{p+q}+a_{-(p+q)}a_pa_q),
\end{align}
where the second equality follows from a change of indices and the commutation relation Eq.~(\ref{eqt:com}).

In this paper, we apply the non-degenerate perturbation method directly and explain its validity at the end of this subsection. According to the results,
\begin{equation}
|\{{n_k}\}^{(1)}\rangle=-\sum_{\{n_k'\}\neq \{n_k\}}\frac{\langle\{{n_k'}\}^{(0)}|D^{(1)}|\{n_k\}^{(0)}\rangle}{d^{(0)}(\{n_k'\})-d^{(0)}(\{n_k\})}|\{{n_k'}\}^{(0)}\rangle .
\end{equation}
Two types of terms of $D$ make contributions: $a_{-p}a_{-q}a_{p+q}$, which splits one occupation on the $p+q$ state into one on $p$ and another on $q$; and $a_{-(p+q)}a_pa_q$, which annihilates one occupation on $p$ and one on $q$, and creates one on $p+q$. For each state obtained by acting with $a_{-p}a_{-q}a_{p+q}$ or $a_{-(p+q)}a_pa_q$ on the ket $|\{n_k\}^{(0)}\rangle$, the product with the bra $\langle\{{n_k'}\}^{(0)}|$ is nonzero only if the two represent the same state up to an overall factor, whose condition can be derived with Eq.~(\ref{eqt:creann}). Therefore, only two kinds of channels make contributions to the expression of the first-order correction: splitting channels and joining channels. One denotes
\begin{equation}
\begin{cases}
\displaystyle|{\psi_{spq}}^{(0)}\rangle=|\cdots,\,n_p+1,\,\cdots,\,n_q+1,\,\cdots,\,n_{p+q}-1,\,\cdots\rangle_{\mathrm{Fock}},\\
\displaystyle|{\psi_{jpq}}^{(0)}\rangle=|\cdots,\,n_p-1,\,\cdots,\,n_q-1,\,\cdots,\,n_{p+q}+1,\,\cdots\rangle_{\mathrm{Fock}}.
\end{cases}
\end{equation}
Interchanging $(p,\,q)$ gives the same channel, because the total operator corresponds to the same final state, so there are exactly four kinds of channels: $p<q$ splitting, $p=q$ splitting, $p<q$ joining, and $p=q$ joining. Finally, the first-order correction of the eigenstates is written as
\begin{equation}\label{eqt:eigcor}
\begin{aligned}
&\,|\{n_k\}^{(1)}\rangle\\
=&\,\sum_{0<p<q} \sqrt{h}\sqrt{\frac{(n_p+1)(n_q+1)n_{p+q}}{pq(p+q)}}|{\psi_{spq}}^{(0)}\rangle+\sum_{p>0} \frac{\sqrt{h}}{4}\sqrt{\frac{(n_p+1)(n_p+2)n_{2p}}{2p^3}}|{\psi_{spp}}^{(0)}\rangle\\
&\,-\sum_{0<p<q} \sqrt{h}\sqrt{\frac{n_pn_q(n_{p+q}+1)}{pq(p+q)}}|{\psi_{jpq}}^{(0)}\rangle-\sum_{p>0} \frac{\sqrt{h}}{4}\sqrt{\frac{n_p(n_p-1)(n_{2p}+1)}{2p^3}}|{\psi_{jpp}}^{(0)}\rangle.
\end{aligned}
\end{equation}

Although there are still some accidental degeneracies with only $Q_1$ and $Q_3$, it is appropriate to apply the non-degenerate perturbation method, by virtue of Theorem \ref{theo:1} together with the first-order diagonal condition
\[
\langle\{n_k'\}^{(0)}|D^{(1)}|\{n_k\}^{(0)}\rangle=0,
\]
where $\{n_k'\}\neq\{n_k\}$ in the degenerate subspace with respect to $Q_1$ and $D^{(0)}$, because the splitting and joining channels do not connect two degenerate states, referring to Eq.~(\ref{eqt:l3d}) and the parity condition illustrated in Section \ref{sec:genstr}. Furthermore, the second-order diagonal condition
\begin{equation}\label{eqt:addition}
\langle\{n_k'\}^{(0)}|D^{(2)}|\{n_k\}^{(0)}\rangle-\sum_{\{n_k''\}\neq\{n_k\},\,\{n_k'\}} \frac{\langle\{n_k'\}^{(0)}|D^{(1)}|\{n_k''\}^{(0)}\rangle\langle\{n_k''\}^{(0)}|D^{(1)}|\{n_k\}^{(0)}\rangle}{d^{(0)}(\{n_k''\})-d^{(0)}(\{n_k\})}=0,
\end{equation}
where $\{n_k'\}\neq\{n_k\}$ in the degenerate subspace, should also be taken into account, because there is also $\langle\{n_k\}^{(0)}|D^{(1)}|\{n_k\}^{(0)}\rangle=0$. All $|\{n_k'\}\rangle$'s degenerating with $|\{n_k\}\rangle$ with respect to $Q_1$ and the leading order of $D^{(0)}$ share the same $M_1$ and $M_3$. A calculation shows that, for such $|\{n_k'\}\rangle$, Eq.~(\ref{eqt:addition}) is satisfied. By calculating for $\{n_k'\}=\{n_k\}$, one can even derive
\[
\ell^4 q_3=\tilde{\Delta}^2+(6M_1\tilde{\Delta}+4M_3\tilde{c})+\left[M_3+3M_1^2+\frac{3\tilde{\Delta}}{2\tilde{c}}(2M_1-M_{0,\,2})\right]+O(\tilde{c}^{-1}),
\]
which fits with the result in Ref.~\cite{Dymarsky:2022dhi} in the thermodynamic limit. The specific procedures are the same as the calculations in Section \ref{sec:6}.

If the degeneracy persists when the calculation is carried to the corresponding order, one then goes two orders higher, repeating this process until the degeneracy is eventually removed. According to the discussion in Section \ref{sec:genstr}, together with Theorem \ref{theo:1}, the infinitely-perturbative correction obtained in this way is the exact simultaneous eigenstate of all the qKdV charges. One can substitute the solved eigenstates into the expectation value of a qKdV charge, which is a general method of calculating subleading orders of the qKdV charges' eigenvalues. Plugging the simultaneous eigenstates of $Q_1$ and $Q_3$ solved above into $\langle Q_5\rangle$, one can reproduce the result Eq.~(4.22) of Ref.~\cite{Dymarsky:2022dhi}.
\section{Expectation values of vacuum-family quasi-primaries}\label{sec:expectations}

The dependence of the qKdV charges and of the quasi-primary expectation values on the linear occupation-number moments $\{M_{2r-1}\}$ further indicates that the expectation of any quasi-primary in the vacuum family $\langle \mathcal{O}_{2a}^{(b)}\rangle$, where $2a$ is the level and $b$ is the quasi-primary ordering index, can only be expressed in terms of qKdV charges with indices lower than $2a$ --- this is also the original form of truncated GETH~\cite{Dymarsky:2019etq} --- because those with higher indices introduce occupation-number moments of their own indices through Eq.~(\ref{eqt:m2q}), which the quasi-primary expectation values do not possess; meanwhile, the new structure cannot be expressed in terms of the occupation-number moments appearing in the expectation values, where this independence can be proved with the Jacobian analysis, as can also be repeatedly seen in Section \ref{sec:rgeth}. Truncated GETH holds for vacuum-family quasi-primaries at levels 2 and 4, as the energy-momentum tensor $T=q_1$ and $\mathcal{A}\mathrel{:=}(TT)-\frac{3}{10}\partial^2T=q_3$, so this section focuses on levels 6 and 8, aiming to reveal the general structure. We follow Ref.~\cite{DiFrancesco:1997nk} and use $(\cdots)$ to denote normal ordering. 

Quasi-primaries in the vacuum family have general expressions. The terms giving the leading-order contribution in $c$ after taking the expectation value can be taken to be those normally ordered products of the energy-momentum tensor and its derivatives, ordered from left to right by derivative order, in which the highest-order derivative occurs at least twice, with all remaining derivatives having arbitrary orders no greater than the highest one. By applying the projection operator
\begin{equation}
\Pi_N=\sum_{m=0}^{N-2}(-1)^m\frac{(2N-m-2)!}{m!\,(2N-2)!}L_{-1}^m L_1^m,
\end{equation}
in which $N$ is the level, to the states corresponding to these kernels~\cite{Tolstoy:2004tp}, one obtains the linear space spanned by the corresponding quasi-primaries by reversing the operator--state correspondence. This is a general approach assisted by AI; however, this section adopts another convention that is closer to what is most commonly used in the literature~\cite{Chen:2013dxa,Dymarsky:2019etq}. Generally, the $\ell$-scaling of a quasi-primary expectation $\langle\mathcal{O}_{2a}^{(b)}\rangle$ is $O(\ell^{2a})$~\cite{Dymarsky:2022dhi}.
\subsection{Level 6}\label{sec:6}

At level 6, there are two quasi-primaries in the vacuum family~\cite{Dymarsky:2019etq}
\begin{numcases}{}
\mathcal{O}_6^{(1)}=(T(TT))-\frac{9}{10}(T\partial^2T)+\frac{4}{35}\partial^4T,\\
\mathcal{O}_6^{(2)}=(\partial T\,\partial T)-\frac{4}{5}(T\partial^2T)+\frac{23}{210}\partial^4T.
\end{numcases}
For simplicity, we remove the terms that are introduced to force the orthogonality. By translational invariance, the one-point function of a total derivative vanishes in a momentum eigenstate~\cite{Dymarsky:2019iny}. Exploiting the freedom to add total derivatives, one derives
\begin{numcases}{}
\langle\mathcal{O}_6^{(1)}\rangle=\langle(T(TT))\rangle+\frac{9}{10}\langle(\partial T\,\partial T)\rangle,\\
\langle\mathcal{O}_6^{(2)}\rangle=\frac{9}{5}\langle(\partial T\,\partial T)\rangle.
\end{numcases}

The key to expressing the quasi-primary expectation values in terms of qKdV charges is to express $\langle(\partial T\,\partial T)\rangle$ in terms of the latter. In the thermodynamic limit,
\begin{equation}
\langle(\partial T\,\partial T)\rangle=2\sum_{k=1}^{\infty}k^2 \langle L_{-k} L_k\rangle=2\left\langle \sum_{k=1}^{\infty}k^2 L_{-k} L_k\right\rangle .
\end{equation}

This subsection now focuses on the expression of $\langle \sum_{k=1}^{\infty}k^2 L_{-k} L_k\rangle$. Denote $ \sum_{k=1}^{\infty}k^2 L_{-k} L_k$ by $\mathcal{D}$. The expectation value up to $O(\tilde{c}^0)$ is
\begin{equation}
\begin{aligned}
\langle \mathcal{D} \rangle=&\ \tilde{c}\,\langle\{{n_k}\}^{(0)}|\mathcal{D}^{(0)}|\{n_k\}^{(0)}\rangle+\langle\{{n_k}\}^{(1)}|\mathcal{D}^{(0)}|\{n_k\}^{(1)}\rangle+\langle\{{n_k}\}^{(1)}|\mathcal{D}^{(1)}|\{n_k\}^{(0)}\rangle\\
&+\langle\{{n_k}\}^{(0)}|\mathcal{D}^{(1)}|\{n_k\}^{(1)}\rangle+\langle\{{n_k}\}^{(0)}|\mathcal{D}^{(2)}|\{n_k\}^{(0)}\rangle+\langle\{{n_k}\}^{(2)}|\mathcal{D}^{(0)}|\{n_k\}^{(0)}\rangle\\
&+\langle\{{n_k}\}^{(0)}|\mathcal{D}^{(0)}|\{n_k\}^{(2)}\rangle+O(\tilde{c}^{-1}).
\end{aligned}
\end{equation}
This is because the odd-order corrections of $\langle \mathcal{D} \rangle$ respectively sum to zero. Specifically, one reads from Eq.~(\ref{eqt:rea}) that
\begin{numcases}{}
\mathcal{D}^{(0)}=2\sum_{k=1}^{\infty}k^2(k^2+h)a_{-k}a_k,\\
\mathcal{D}^{(1)}=\sqrt{2}\sum_{k=0}^{\infty}k^2\left[\left(\sqrt{h}-ik\right)a_{-k}\left(\frac{1}{2}\sum_{p=1}^{k-1}a_pa_{k-p}+\sum_{p=1}^{\infty}a_{-p}a_{k+p}\right)\right.\nonumber\\
\phantom{\mathcal{D}^{(1)}=}\left.+\left(\sqrt{h}+ik\right)a_{-k}\left(\frac{1}{2}\sum_{p=1}^{k-1}a_{-p}a_{-(k-p)}+\sum_{p=1}^{\infty}a_{-(k+p)}a_{p}\right)\right],\label{eqt:1cor}\\
\mathcal{D}^{(2)}=\left(\frac{1}{2}\sum_{p=1}^{k-1}a_pa_{k-p}+\sum_{p=1}^{\infty}a_{-p}a_{k+p}\right)^\dagger\left(\frac{1}{2}\sum_{p=1}^{k-1}a_pa_{k-p}+\sum_{p=1}^{\infty}a_{-p}a_{k+p}\right).\label{eqt:2cor}
\end{numcases}

Combining the normalization condition with the result that $|\{n_k\}^{(0)}\rangle$ is an eigenstate of $\mathcal{D}$ gives
\begin{equation}
\begin{aligned}
&\langle\{{n_k}\}^{(1)}|\mathcal{D}^{(0)}|\{n_k\}^{(1)}\rangle+\langle\{{n_k}\}^{(2)}|\mathcal{D}^{(0)}|\{n_k\}^{(0)}\rangle+\langle\{{n_k}\}^{(0)}|\mathcal{D}^{(0)}|\{n_k\}^{(2)}\rangle\\
=&\sum_{\{n_k'\}\neq \{n_k\}}\left|\frac{\langle\{{n_k'}\}^{(0)}|D^{(1)}|\{n_k\}^{(0)}\rangle}{d^{(0)}(\{n_k'\})-d^{(0)}(\{n_k\})}\right|^2\left(E^{(0)}(\{n_k'\})-E^{(0)}(\{n_k\})\right).
\end{aligned}
\end{equation}
Here, $E$ is used to denote the eigenvalues of $\mathcal{D}$, not the energy. The channels are exactly the same as those in Section \ref{sec:cor}.
\begin{table}[htbp]
\centering
\begin{tabular}{c|cc}
 & $\left|\frac{\langle\{{n_k'}\}^{(0)}|D^{(1)}|\{n_k\}^{(0)}\rangle}{d^{(0)}(\{n_k'\})-d^{(0)}(\{n_k\})}\right|^2$& $E^{(0)}(\{n_k'\})-E^{(0)}(\{n_k\})$ \\
\hline
$p<q$ splitting & $\frac{h}{2pq(p+q)}(n_p+1)(n_q+1)n_{p+q}$ & $-2pq(p+q)[3h+5(p^2+pq+q^2)]$ \\
$p<q$ joining & $\frac{h}{2pq(p+q)}n_pn_q(n_{p+q}+1)$ & $+2pq(p+q)[3h+5(p^2+pq+q^2)]$ \\
$p=q$ splitting & $\frac{h}{16p^3}(n_p+1)(n_p+2)n_{2p}$ & $-12p^3(h+5p^2)$ \\
$p=q$ joining & $\frac{h}{16p^3}n_p(n_p-1)(n_{2p}+1)$ & $+12p^3(h+5p^2)$
\end{tabular}
\caption{Calculation of each kind of channel.}\label{tab:channel}
\end{table}

For each of the four rows in Table \ref{tab:channel}, we first sum the two quantities in the left and right columns. For each $(p,\,q)$, the results from $p<q$ and $p=q$ are then summed separately, followed by summation over their respective indices, yielding
\[
-h\sum_{0<p<q}[3h+5(p^2+pq+q^2)][n_{p+q}(n_p+n_q+1)-n_pn_q]-\frac{3h}{4}\sum_{p>0}(h+5p^2)(4n_pn_{2p}+2n_{2p}-n_p^2+n_p).
\]
After relabeling the indices to simplify this expression, all terms that cannot be written in the form of $M_{p,\,s}$ cancel out, leaving
\begin{equation}
\begin{aligned}
&\langle\{{n_k}\}^{(1)}|\mathcal{D}^{(0)}|\{n_k\}^{(1)}\rangle+\langle\{{n_k}\}^{(2)}|\mathcal{D}^{(0)}|\{n_k\}^{(0)}\rangle+\langle\{{n_k}\}^{(0)}|\mathcal{D}^{(0)}|\{n_k\}^{(2)}\rangle\\
=&\ 5hM_1^2-\frac{5}{4}hM_{2,\,2}+\frac{3}{4}h^2M_{0,\,2}-\frac{25}{12}hM_3-\frac{5}{4}hM_2-\frac{5}{12}hM_1-\frac{3}{2}h^2M_1+\frac{3}{4}h^2M_0.
\end{aligned}
\end{equation}
And with Eq.~(\ref{eqt:2cor}), we get
\begin{equation}
\langle\{{n_k}\}^{(0)}|\mathcal{D}^{(2)}|\{n_k\}^{(0)}\rangle=2M_1M_3-M_{4,\,2}+\frac{1}{12}M_5-M_4-\frac{1}{12}M_3 .
\end{equation}

Finally, we consider the remaining terms $\langle\{{n_k}\}^{(1)}|\mathcal{D}^{(1)}|\{n_k\}^{(0)}\rangle+\langle\{{n_k}\}^{(0)}|\mathcal{D}^{(1)}|\{n_k\}^{(1)}\rangle$. Indeed,
\begin{equation}
\langle\{{n_k}\}^{(1)}|\mathcal{D}^{(1)}|\{n_k\}^{(0)}\rangle=\sum_{\{n_k'\}\neq \{n_k\}}\left[-\frac{\langle\{n_k\}^{(0)}|D^{(1)}|\{{n_k'}\}^{(0)}\rangle}{d^{(0)}(\{n_k'\})-d^{(0)}(\{n_k\})}\right]\langle\{{n_k'}\}^{(0)}|\mathcal{D}^{(1)}|\{n_k\}^{(0)}\rangle .
\end{equation}
For convenience, one considers the contributions to $\mathcal{D}^{(1)}$ separately for each $k$. For $p<q$ splitting channels, only three terms of $D^{(1)}$ contribute to nonzero terms:
\begin{itemize}
\item $k=p$, $(L_{-k}L_k)^{(1)}\supset \sqrt{2} (\sqrt{h}-ip) a_{-p}a_{-q}a_{p+q}$.
\item $k=q$, $(L_{-k}L_k)^{(1)}\supset \sqrt{2} (\sqrt{h}-iq) a_{-p}a_{-q}a_{p+q}$.
\item $k=p+q$, $(L_{-k}L_k)^{(1)}\supset \sqrt{2} (\sqrt{h}+i(p+q)) a_{-p}a_{-q}a_{p+q}$.
\end{itemize}
After summing over $k$, the total contribution from this kind of channel and its Hermitian conjugate is
\[
\sum_{k=1}^{\infty} k^2\sum_{p<q}2h(n_p+1)(n_q+1)n_{p+q}(\delta_{kp}+\delta_{kq}+\delta_{k,\,p+q}).
\]
Applying the same procedure to the other three classes of channels and performing the corresponding algebraic manipulations, one obtains
\begin{equation}
\langle\{{n_k}\}^{(1)}|\mathcal{D}^{(1)}|\{n_k\}^{(0)}\rangle+\langle\{{n_k}\}^{(0)}|\mathcal{D}^{(1)}|\{n_k\}^{(1)}\rangle=-4hM_1^2+hM_{2,\,2}+\frac{5}{3}hM_3+hM_2+\frac{1}{3}hM_1 .
\end{equation}
After taking the thermodynamic limit Eq.~(\ref{eqt:llim}),
\begin{equation}\label{eqt:add}
\begin{aligned}
\langle \mathcal{D} \rangle=&\ 2(\tilde{\Delta}M_3+\tilde{c}M_5)+\left(2M_1M_3+\frac{1}{12}M_5\right)+\frac{\tilde{\Delta}}{\tilde{c}}\left(M_1^2-\frac{5}{12}M_3\right)\\
&+\frac{3\tilde{\Delta}^2}{4\tilde{c}^2}\left(M_{0,\,2}-2M_1\right)+O(\tilde{c}^{-1}).
\end{aligned}
\end{equation}

Considering the relationship between $\langle \mathcal{O}_6^{(2)}\rangle$ and $\langle \mathcal{D} \rangle$, with Eqs.~(\ref{eqt:dq3}) and (\ref{eqt:dq5}), after simplification, Eq.~(\ref{eqt:add}) leads to
\begin{equation}\label{eqt:o62}
q_1^{-3}\langle \mathcal{O}_6^{(2)}\rangle=\frac{9}{10\tilde{c}}(\delta q_5-3\delta q_3)-\frac{9}{80\tilde{c}^2}(5\delta q_5+9\delta q_3)+\frac{27}{5\tilde{\Delta}\tilde{c}^2}M_{0,\,2}+O(\tilde{c}^{-4}) .
\end{equation}
Note that $q_5=\mathcal{O}_6^{(1)}-\frac{93}{70c+29}\mathcal{O}_6^{(2)}+\frac{5c}{108}\mathcal{O}_6^{(2)}$~\cite{Dymarsky:2019etq}, from which we derive that
\[
q_1^{-3}\langle \mathcal{O}_6^{(1)}\rangle=1+\delta q_5+\left(\frac{93}{70c+29}-\frac{5c}{108}\right)q_1^{-3}\langle \mathcal{O}_6^{(2)}\rangle,
\]
leading to
\begin{equation}
q_1^{-3}\langle \mathcal{O}_6^{(1)}\rangle=1+3\delta q_3+\frac{1}{12\tilde{c}}(7\delta q_5+15\delta q_3)-\frac{6}{\tilde\Delta\tilde c}M_{0,\,2}+O(\tilde c^{-3}) .
\end{equation}

The expectation values fail to be expressed as functions of a finite number of the first qKdV charges, as $M_{0,\,2}$ is not a function of $(q_1,\,q_3,\,q_5)$, whose proof is rigorously given in Section \ref{sec:rgeth}. Nevertheless, there is evidence that the subleading orders of quasi-primary expectation values can be regarded as functionals of $\left\{q_{2k-1}(\tilde{c})\right\}$ according to the previous calculations. Expand $\{\delta q_{2k-1}(\tilde c)\}\ (k\geq2)$ by the orders of $\tilde c$:
\begin{equation}
\delta q_{2k-1}=\frac{1}{\tilde{c}}\delta q_{2k-1}^{(1)}+\frac{1}{\tilde{c}^2}\delta q_{2k-1}^{(2)}+O(\tilde{c}^{-3}),
\end{equation}
and then $M_{0,\,2}$ can be expressed by $q_3(\tilde{c}),\,q_5(\tilde{c})$, having substituted $\left(m,\,\delta q_3^{(1)},\,\delta q_5^{(1)}\right)$ into \hfill\break $\left(\delta q_3^{(2)},\,\delta q_5^{(2)}\right)$, eliminated $M_{0,\,2}$ temporarily, solved the quadratic equation for $m$, retained the physical root, and plugged it back:
\begin{equation}\label{eqt:m02}
M_{0,\,2}=\frac{h}{96}\left[A+\left(7\delta {q_3^{(1)}}-3\delta {q_5^{(1)}}\right)\sqrt{B}\right]
\end{equation}
where
\begin{numcases}{\quad\quad}
A=9\delta {q_5^{(1)}}^2+105\delta {q_3^{(1)}}^2-66\delta {q_3^{(1)}}\delta {q_5^{(1)}}-16\delta {q_5^{(2)}}+16\delta {q_3^{(2)}}+\frac{104}{3}\delta {q_5^{(1)}}-40\delta q_3^{(1)},\nonumber\\
B=9\delta {q_5^{(1)}}^2+225\delta {q_3^{(1)}}^2-90\delta {q_3^{(1)}}\delta {q_5^{(1)}}-32\delta {q_5^{(2)}}+160\delta {q_3^{(2)}}+\frac{208}{3}\delta {q_5^{(1)}}-240\delta q_3^{(1)}+64.\nonumber
\end{numcases}
Equivalently, $M_{0,\,2}$ is expressible up to the leading order, and then $\langle\mathcal{O}_6^{(1)}\rangle$ and $\langle\mathcal{O}_6^{(2)}\rangle$ are expressible up to the subleading order by $\delta q_3,\,\delta q_5,\,\partial_{\tilde{c}}\delta q_3,\,\partial_{\tilde{c}}\delta q_5$, with $h$ seen as fixed.

\subsection{Level 8}\label{sec:8}

At level 8, there are three quasi-primaries in the vacuum family~\cite{Chen:2013dxa}
\begin{numcases}{}
\mathcal{O}_8^{(1)}=(T(T(TT))) - \frac{9}{5}(T(T \partial^2 T)) + \frac{3}{10}(\partial T \partial^3 T) + \frac{29}{70}(T \partial^4 T) - \frac{41}{1680}\partial^6 T,\\
\mathcal{O}_8^{(2)}=(T(\partial T \partial T)) - \frac{4}{5}(T(T \partial^2 T)) - \frac{1}{5}(\partial T \partial^3 T) + \frac{47}{210}(T \partial^4 T) - \frac{29}{2520}\partial^6 T,\\
\mathcal{O}_8^{(3)}=(\partial^2 T \partial^2 T) - \frac{10}{9}(\partial T \partial^3 T) + \frac{10}{63}(T \partial^4 T) - \frac{13}{2268}\partial^6 T.
\end{numcases}
For the process using the vanishing of the expectation value of total-derivative terms, one may refer to the formulae in Ref.~\cite[Appendix 6.C]{DiFrancesco:1997nk}. The results are
\begin{numcases}{}
\langle\mathcal{O}_8^{(1)}\rangle=\langle (T(T(TT))) \rangle + \frac{18}{5} \langle (T(\partial T \partial T)) \rangle + \frac{29}{70} \langle (\partial^2 T \partial^2 T) \rangle,\\
\langle\mathcal{O}_8^{(2)}\rangle=\frac{13}{5} \langle (T(\partial T \partial T)) \rangle + \frac{39}{70} \langle (\partial^2 T \partial^2 T) \rangle,\\
\langle\mathcal{O}_8^{(3)}\rangle=\frac{143}{63}\langle(\partial^2T\,\partial^2T)\rangle.
\end{numcases}

In the thermodynamic limit one conducts transformations and gets~\cite{Dymarsky:2019iny}
\begin{numcases}{}
\langle(\partial^2T\,\partial^2T)\rangle=&$\displaystyle2\sum_{k=1}^{\infty}k^4 \langle L_{-k} L_k\rangle,$\\
\langle (T(\partial T \partial T)) \rangle=&$\displaystyle\sum_{k,\,l=1}^{\infty} (k+l)l\ \langle L_{-k} L_{-l} L_{k+l} \rangle+ \sum_{l=0,\,k=1}^{\infty} (k-l)k\ \langle L_{-k} L_{k-l} L_l\rangle$\nonumber\\
&$\displaystyle+\sum_{k,\,l=0}^{\infty} (k+l)k\ \langle L_{-k-l} L_k L_l\rangle+ \sum_{k=0,\,l=1}^{\infty} (k-l)k\ \langle L_{-l} L_{l-k} L_k\rangle.$\label{eqt:lll}
\end{numcases}
For these two quantities, one adopts the same treatment as for that of level 6, converting the calculation of correlation functions into the computation of creation and annihilation operators acting on free-bosonic states, channel by channel.

Repeating the formulation in Section \ref{sec:6} gives
\begin{equation}\label{eqt:1}
\begin{aligned}
\langle(\partial^2T\,\partial^2T)\rangle=&\ 2\left[(2M_5\tilde{\Delta}+2\tilde{c}M_7)+\left(2M_1 M_5+6M_3^2+\frac{1}{30}M_7\right)\phantom{+\frac{5\tilde{\Delta}^2}{\tilde{c}^2}}\right.\\&\ \left.+\frac{\tilde{\Delta}}{\tilde{c}}\left(12M_1M_3-\frac{21}{20}M_5\right)+\frac{5\tilde{\Delta}^2}{\tilde{c}^2}\left(M_1^2-\frac{5}{12}M_3\right)\right]+O(\tilde{c}^{-1})
\end{aligned}
\end{equation}
in the thermodynamic limit, and then the subleading order of $\langle\mathcal{O}_8^{(3)}\rangle$ can be reduced to qKdV charges:
\begin{equation}
\begin{aligned}
q_1^{-4}\langle\mathcal{O}_8^{(3)}\rangle=&\ \frac{143}{63}\left[\frac{5}{16\tilde{c}^2}\left(\delta q_7-4\delta q_5+6\delta q_3\right)+\frac{5}{16\tilde{c}^2}(\delta q_3)^2\right.\\&\left.-\frac{5}{192\tilde{c}^3}\left(14\delta q_7-4\delta q_5-72\delta q_3\right)\right]+O(\tilde c^{-5}).
\end{aligned}
\end{equation}

Next one observes the scalings of different parts of Eq.~(\ref{eqt:lll}), where Eq.~(\ref{eqt:lll}) can be split into the boundary part
\[
2\sum_{k=1}^{\infty}k^2 \langle L_{-k}L_kL_0\rangle
\]
and the bulk part
\[
\begin{aligned}
&\ \sum_{k,\,l=1}^{\infty} (k+l)l\ \langle L_{-k} L_{-l} L_{k+l} \rangle+ \sum_{k,\,l=1}^{\infty} (k-l)k\ \langle L_{-k} L_{k-l} L_l\rangle\\
+&\ \sum_{k,\,l=1}^{\infty} (k+l)k\ \langle L_{-k-l} L_k L_l\rangle+ \sum_{k,\,l=1}^{\infty} (k-l)k\ \langle L_{-l} L_{l-k} L_k\rangle.
\end{aligned}
\]
One needs to consider the leading and subleading orders of the boundary part and only the leading order of the bulk part according to their scalings. The boundary part is not hard to obtain, since the states under discussion are $L_0$-eigenstates:
\[
2(\tilde{\Delta}+M_1)\langle D\rangle,
\]
while the bulk part is constituted of two subparts, both of which are at the $\tilde c$-scaling $O(\tilde{c}^2)$. The first one is contributed by two leading orders of $L$ and one subleading order of it acting on the zeroth-order Fock states, since $O(\tilde{c}^{\frac{3}{2}})$ vanishes, which is derived in the thermodynamic limit as
\[
\left(16M_1 M_3+\frac{7}{3}M_5\right)\tilde{\Delta}+\left(-8M_3^2+\frac{9}{5}M_7\right)\tilde{c}+O(\tilde{c}^{0}).
\]
And the second subpart is contributed by three leading orders of $L$ acting on a leading and a subleading order of the eigenstate:
\[
\left(\frac{14}{5}M_5-32M_1M_3\right)\tilde{\Delta}+\left(\frac{10}{3}M_3-8M_1^2\right)\frac{\tilde{\Delta}^2}{\tilde{c}}+O(\tilde{c}^{0}).
\]
Add these three together and finally, in the thermodynamic limit,
\begin{align}
\langle (T(\partial T \partial T)) \rangle=&\ 4\tilde{\Delta}(\tilde{\Delta}M_3+\tilde{c}M_5)+\tilde{\Delta}\left(\frac{53}{10}M_5-8M_1M_3\right)+\tilde{c}\left(4M_1M_5-8M_3^2+\frac{9}{5}M_7\right)\nonumber\\
&+\frac{2\tilde{\Delta}^2}{\tilde{c}^2}\left(\frac{5}{4}M_3-3M_1^2\right)+\frac{3\tilde{\Delta}^3}{2\tilde{c}^2}(M_{0,\,2}-2M_1)+O(\tilde{c}^{0}).
\end{align}
After simplification, one obtains
\begin{equation}\label{eqt:2}
\begin{aligned}
q_1^{-4}\left\langle (T(\partial T\partial T))\right\rangle=&\ \frac{1}{2\tilde c}\left(\delta q_5-3\delta q_3\right)-\frac{1}{2\tilde c}\left(\delta q_3\right)^2+\frac{1}{192\tilde{c}^2}\left(27\delta q_7-88\delta q_5-186\delta q_3\right)\\
&\ +\frac{3}{\tilde{\Delta}\tilde c^2}M_{0,\,2}+O(\tilde c^{-4}).
\end{aligned}
\end{equation}
Combining Eqs.~(\ref{eqt:1}) and (\ref{eqt:2}), one obtains
\begin{equation}\label{eqt:o82}
\begin{aligned}
q_1^{-4}\langle\mathcal{O}_8^{(2)}\rangle=&\ \frac{13}{10\tilde{c}}\left[(\delta q_5-3\delta q_3)-(\delta q_3)^2+\frac{1}{672\tilde{c}}(279\delta q_7-976\delta q_5-762\delta q_3)+\frac{6}{\tilde{\Delta}\tilde{c}}M_{0,\,2}\right]\\&\ +O(\tilde{c}^{-4}).
\end{aligned}
\end{equation}

Indeed, qKdV charge operators are able to be expressed in terms of vacuum-family quasi-primaries at the corresponding levels~\cite{Dymarsky:2019iny}, which is related to the convention where orthogonality is restored, and here readers are referred to Ref.~\cite[Appendix A]{Chen:2013dxa} for details. Hence, all possible non-qKdV-charge structures in $\langle\mathcal{O}_8^{(1)}\rangle$ are introduced by those in $\langle\mathcal{O}_8^{(2)}\rangle$ and $\langle\mathcal{O}_8^{(3)}\rangle$, so one does not necessarily calculate $\langle\mathcal{O}_8^{(1)}\rangle$ for the discussion. Like $\langle\mathcal{O}_6^{(2)}\rangle$, $\langle\mathcal{O}_8^{(1)}\rangle$ and $\langle\mathcal{O}_8^{(2)}\rangle$ are functionals of qKdV charges and their derivatives with respect to $\tilde{c}$, from Eq.~(\ref{eqt:m02}).

As a note, the attempt to express the expectation values of quasi-primary operators $\{\langle\mathcal{O}_{2a}^{(b)}\rangle\}$ as functions of the qKdV charges in Sections \ref{sec:6} and \ref{sec:8} consists of three steps. In the first step, the leading and subleading orders are expressed in terms of occupation-number moments; specifically, $\{M_{2r-1}\}$ with $r\leq a$, and $M_{0,\,2}$ for those at levels 6 and 8. The manipulations at both orders are exact identities, without omission of $\tilde{c}$- and $\ell$-subleading orders. Then, one takes the thermodynamic limit and finds that most terms that might cause departure from the qKdV charges, like $M_{2r,\,2}$ with $r\geq1$, are omitted. The third step replaces these occupation-number moments using their expressions in terms of the qKdV charges, and there is no explicit $m$ in the final expressions. The replacement performed at an order also affects the next order, where the transformation of the subleading order in terms of the qKdV charges neglects a sub-subleading-order correction. Further calculation of the third order of $\langle\mathcal{O}_8^{(2)}\rangle$ indicates that there are a small proportion of terms that cannot be reduced to occupation-number moments, because $LLL$-form Hermitians are not diagonal like $L_{-k}L_k$, so the numerators do not cancel the denominators out.

\section{Distinguishability by and completeness of qKdV charges}\label{sec:discomp}
This section discusses the problems of whether the qKdV charges distinguish all microstates, and whether GGE should be further extended, including quantities other than qKdV charges. Discussion on these properties needs precise values of qKdV charges, so the thermodynamic limit is not considered in this section.
\subsection{Within a conformal family}\label{sec:conformal}
We first restrict our consideration to a fixed conformal family, where $\Delta$, or equivalently $\tilde{\Delta}$, is regarded as a known constant. First, all $n_k$, i.e.~the occupations of the bosonic states, can be recovered by the complete set of $M_{2r-1}$, as is shown in Appendix \ref{app:reconst}:
\begin{equation}\label{eqt:reconst}
n_k=-2(-1)^k \sum_{r=1}^{\infty}\sum_{m=0}^{r-1}\frac{(-1)^m(\pi k)^{2m}}{(2m+1)!}\frac{M_{2r-1}}{k^{2r-1}} .
\end{equation}
Second, with $\Delta$ given, the leading terms of $\{\delta q_{2k-1}\}\ (k\geq2)$ together with $m$ form a lower triangular system in the moments $\{{M_{2k-1}}\}$ from Eq.~(\ref{eqt:sum}), so the transformation matrix has full rank, allowing these moments to be eliminated recursively in favor of the qKdV charges, order by order, whose $\tilde{c}$-leading-order calculation is shown in Eq.~(\ref{eqt:inversion}). In conclusion, within any fixed conformal family, $n_k$ can be solved by qKdV charges inversely, so the microstates and qKdV charges form a one-to-one correspondence. Equivalently, simultaneous eigenstates of all the qKdV charges are non-degenerate in any fixed conformal family, as is concluded by Theorem \ref{theo:2}, because each simultaneous eigenstate can be distinguished. 
\begin{theorem}\label{theo:2}
For a given $(c,\,|\Delta\rangle)$, the full set of qKdV charges can lift all the degeneracies.
\end{theorem}
That is to say, within a fixed conformal family the set of all qKdV charges is a complete set of commuting quantities (CSCO). 

It demonstrates that the reconstruction from the infinite tower is informationally possible; however, it does not admit a rapidly convergent infinite series in the thermodynamic limit. In a computational sense, this requires the technique of analytic continuation. The full set of qKdV charges defines, according to the above rule, an analytic function in $\{q_{2k-1}\}$; one then needs to determine this function within a region of rapid convergence of the series with respect to $\{q_{2k-1}\}$ and analytically continue it to the scaling under discussion, whose treatment is shown in Appendix \ref{app:inversion}.
\subsection{Within a physical system}\label{sec:physical}
An actual conformal physical system should allow all possible states $|\Delta\rangle$'s additionally characterized by highest weights $\Delta$'s, and the density of primary states at large $\Delta$ is given by~\cite{Cardy:1986ie}
\begin{equation}\label{eqt:density}
g(\Delta)=\mathrm{e}^{\pi\sqrt{\frac{2}{3}(c-1)\Delta}}
\end{equation}
due to modular invariance. Therefore, $m$ and $h$ are not good quantities in the expectations of quasi-primaries, and GETH claims that they should vanish in the thermodynamic limit. 

From the analysis in Section \ref{sec:conformal}, 
\begin{equation}
n_k=n_k(\tilde{\Delta},\,\{q_{2r-1}\}),
\end{equation}
as $\Delta$ is no longer seen as a constant, and the scale $\ell$ is given. Here,~\cite{Dymarsky:2022dhi}
\begin{equation}\label{eqt:tri}
M_1=\ell^2q_1-\tilde{\Delta}+\frac{1}{24},\quad h=\frac{\tilde{\Delta}}{\tilde c}. 
\end{equation}
With the process given in Appendix \ref{app:inversion}, the result is 
\begin{equation}\label{eqt:inversion}
\begin{aligned}
n_k&=-2(-1)^k\sum_{r=1}^{\infty}\frac{1}{k^{2r-1}}\left[\sum_{m=0}^{r-1}\frac{(-1)^m(\pi k)^{2m}}{(2m+1)!}\right]\left[\left(-\frac{\tilde{\Delta}}{\tilde{c}}\right)^{r-1}\left(\ell^2q_1-\tilde\Delta\right)\right.\\
&\phantom{=}\ \left.-\frac12\zeta(1-2r)+2\tilde{c}\,\binom{2r}{r}\left(\frac{\tilde{\Delta}}{4\tilde{c}}\right)^r\sum_{j=2}^{r}(-1)^{r-j}\binom{r}{j}\frac{\delta q_{2j-1}}{2j-1}\right]+O(\tilde{c}^{-1}).
\end{aligned}
\end{equation}
Obviously, $\tilde{\Delta}$ remains in the expression, so $n_k$ is a function of qKdV charges and the highest weight. 

One finds that even all qKdV charges may not distinguish different eigenstates in a real system, because the eigenstate can be uniquely inversely solved for any conformal family subject to the constraint mentioned in Appendix \ref{app:constraint}. Explicit counterexamples of distinguishability are found in the case when two conformal families share the same highest weight. In conclusion, qKdV charges do not form a CSCO in some cases. 

However, the establishment of a CSCO extended from qKdV charges is not hard, as we may consider a formal operator $\mathfrak D$ whose eigenvalue is the same within a conformal family but different across different conformal families. It is certain that $\mathfrak D$ is linear. From the density of states formula Eq.~(\ref{eqt:density}), the number of conformal families is countable, so an operator with discrete eigenvalues can characterize the difference. Specially, if any two highest weights are not exactly the same, we can let the eigenvalues of $\mathfrak D$ just be the highest weight. 

Nevertheless, this may come at the expense of the locality of the operators in the set of conserved quantities. Specifically, the set of qKdV charges admits no further vacuum-family observable commuting with all its members~\cite{Zhao:2026xrz}. That's also why qKdV charges are thought to be complete in the GGE, with attempts of extending GGE having to introduce charges in non-vacuum families. 
\section{Generalized Eigenstate Thermalization}\label{sec:geth}
In this section, we discuss the validity of generalized eigenstate thermalization, covering two versions of GETH with regard to infinite-tower and truncated qKdV charges. We argue that the infinite-tower GETH holds for vacuum-family observables. However, it is demonstrated that the central charge $c$, or $\tilde{c}$, should be regarded as a constant instead of a variable in the discussion of GETH, as is discussed in Appendix \ref{app:der}. Therefore, one concludes that $\langle \mathcal{O}_{2a}^{(b)}\rangle$ may not be expressed as a function of $q_1,\,\cdots,\,q_{2a-1}$, rendering the original form of truncated GETH invalid. 
\subsection{GETH with truncated qKdV charges}\label{sec:rgeth}
The discussion on GETH raises the question if the expectation values can be expressed by qKdV charges. Yet, the ultimate goal of GETH is to determine the expectations of local observables without knowing the full microscopic details of the state~\cite{Dymarsky:2019etq}. We therefore seek a more restrictive form of GETH, asking whether an arbitrary local observable expectation can be determined by only finitely many qKdV charges. 

For the expectations of vacuum-family quasi-primaries of levels 6 and 8, the most difficult term to analyze is the non-negligible $M_{0,\,2}$ in the subleading order. Here, we prove that it indeed cannot be expressed as any linear or non-linear function of $q_3$ and $q_5$, i.e.~$M_{0,\,2}$ can never be expressed as the function of $(q_1,\,q_3,\,q_5)$. 

As preparation, parametrizing quantities in terms of $(q_1,\,q_3,\,q_5)$ is equivalent to parametrizing them in terms of $(m,\,\delta q_3,\,\delta q_5)$, since the two sets span the same space of allowed quantities; $M_1$ is a known parameter, because $m=\frac{M_1}{\tilde{\Delta}}$ is temporarily treated as known.

The first step is to prove that all the terms appearing in the leading and subleading orders of $(m,\,\delta q_3,\,\delta q_5)$, ($M_1,\,M_3,\,M_5,\,M_{0,\,2}$), are indeed independent from each other. We calculate a Jacobian that we artificially choose for convenience:
\begin{equation}
\frac{\partial(M_1,\,M_3,\,M_5,\,M_{0,\,2})}{\partial(n_1,\,n_2,\,n_3,\,n_4)}=
\begin{pmatrix}
1&2&3&4\\1&8&27&64\\1&32&243&1024\\2n_1&2n_2&2n_3&2n_4
\end{pmatrix},
\end{equation}
where $n_1,\,n_2,\,n_3,\,n_4$ are real independent parameters and whose determinant is
\begin{equation}
\mathrm{det}\left(\frac{\partial(M_1,\,M_3,\,M_5,\,M_{0,\,2})}{\partial(n_1,\,n_2,\,n_3,\,n_4)}\right)=-1440(14n_1-14n_2+6n_3-n_4).
\end{equation}
It is non-zero unless $(n_1, n_2, n_3, n_4)$ lies on a unique three-dimensional hyperplane. By successively incrementing the subscript of the independent variable by one, one observes that $M_1,\,M_3,\,M_5,\,M_{0,\,2}$ are independent, unless $\{n_k\}$ fall on a special submanifold with very low dimension. Therefore, for a general theory, $M_1,\,M_3,\,M_5,\,M_{0,\,2}$ are genuinely independent, so $M_{0,\,2}$ cannot be expressed as a function of $(M_3,\,M_5)$. This procedure can be generalized to prove that any finite number of occupation-number moments are mutually independent.

As the next step, we calculate the determinant of the Jacobian
\begin{equation}
\left|\frac{\partial(q_3,\,q_5)}{\partial\left(\frac{M_3}{h^2},\,\frac{M_5}{h^3}\right)}\right|=\frac{32}{\tilde{c}^2}+O(\tilde{c}^{-3})\neq0 .
\end{equation}
Therefore, in the representation space of $\delta q_3$ and $\delta q_5$, at leading order at least one of $M_3$ and $M_5$ necessarily appears, however nonlinear the function may be. Thus, $M_{0,\,2}$ cannot be expressed as a function of $(q_1,\,q_3,\,q_5)$.

In Sections \ref{sec:6} and \ref{sec:8}, we found that the subleading orders in $\tilde{c}$ of expectations $\langle\mathcal{O}_{2a}^{(b)}\rangle$ cannot be always explicitly expressed by qKdV charges with the remaining term $M_{0,\,2}$ of thermodynamically leading order in $\ell$, which cannot be omitted. Furthermore, from this subsection, one knows that $M_{0,\,2}$ is not a function of a finite number of qKdV charges. Therefore, some quasi-primaries are not functions of the first few qKdV charges, even when the thermodynamic limit is taken, so the original form of truncated GETH fails at the subleading order in $c$. Other possibilities of truncated GETH are discussed in Section \ref{sec:outcomes}. 
\subsection{GETH with infinite-tower qKdV charges}\label{sec:sgeth}
Here, we directly discuss the validity of the general version of GETH, considering the infinite tower of qKdV charges. In the formula Eq.~(\ref{eqt:inversion}), $\tilde{\Delta}$ remains in the expression. At first glance, the expression appears to involve the factor $\tilde{\Delta}$ other than qKdV charges, seemingly indicating a breakdown of GETH. Nevertheless, in Appendix \ref{app:constraint} we argue that $\tilde{\Delta}$ can be statistically determined by qKdV charges together with the positive conditions of $\{n_k\}$ in the thermodynamic limit. This is a surprising result: countably infinite inequality constraints eliminate a statistical degree of freedom. 

Thus, all terms explicitly depending only on $\{n_k\}$ and $\tilde{\Delta}$, which can be combined to form any expectation value of local observables in the vacuum family, can be reconstructed, satisfying at any order in $\tilde{c}$. The generality of this result follows from the simple structure of quasi-primaries and qKdV charges, as is derived in Section \ref{sec:genstr}. 

A noteworthy point is that the infinite-tower GETH in 2D CFTs is claimed to be satisfied at generic $c$, subject to the inverse problem of chemical potentials, even beyond the large-$c$ limit possibly except for some critical region of degeneration, because the reasoning of Appendix \ref{app:constraint} can be extended to all but special regions of $c$. 
\subsection{Outcomes of GETH}\label{sec:outcomes}
Given that the original truncated GETH fails, we propose three possible prospects for the theory of generalized eigenstate thermalization. 

The first possibility is to seek a more natural basis of qKdV charges. The true reason why GETH without the restoration of microstates fails for $\langle\mathcal{O}_6^{(2)}\rangle$ is the incompatibility between a structure $M_3-\frac{3h}{8\tilde{c}}$ appearing at second and third orders in all qKdV charges~\cite{Dymarsky:2022dhi} and the structure $M_3+\frac{3h}{8\tilde{c}}$ in $\langle\mathcal{O}_6^{(2)}\rangle$; with the disappearance of this problem in $\langle\mathcal{O}_8^{(3)}\rangle$, its subleading order can finally be eliminated using only first four qKdV charges. This suggests that $\{Q_{2k-1}\}$ may not be the most natural basis of the linear space spanned by qKdV charges for some specific problems. A similar attempt of basis transformation can be found in Ref.~\cite{Karimi:2026dnr} for solving another GGE problem. 

The second possibility is to introduce the derivative terms of $q_1,\,\cdots,\,q_{2a-1}$ to express $\langle \mathcal{O}_{2a}^{(b)}\rangle$, which is discussed in Section \ref{sec:6}, and the GGE should be revised as
\begin{equation}
\rho_{\mathrm{GGE}}=\frac{1}{Z_{\mathrm{GGE}}}\mathrm{e}^{-\mathop{\sum_{k=1}^{\infty}}\left( \mu_{2k-1}Q_{2k-1}+\nu_{2k-1}\partial_{\tilde{c}}Q_{2k-1}+\cdots\right)}.
\end{equation}
As a conclusion, this trial is corroborated by the results of the expectations of quasi-primaries of level 6 and level 8, and the expectations at higher levels need to be further discussed to examine its validity. Although it is strange to introduce new quantities other than the chosen set of the conserved charges, the possibility that the quantities are actually linear combinations of the infinite tower of qKdV charges needs to be taken seriously. Broadly speaking, this is another trial of basis transformation just like the first one. 

The third outcome is to seek special points that render $\langle \mathcal{O}_{2a}^{(b)}\rangle$ a function of $q_1,\,\cdots,\,q_{2a-1}$, which may indicate new physics.
\section{Discussion}\label{sec:discussion}

As preparation, we discussed some general features of qKdV charges. We proved first the analytic property of qKdV charges in $\tilde{c}$ and then the non-degenerate property of $Q_1$ and $Q_3$ for any continuous 2D region in $(c,\,\Delta)$. With these preparations, we expressed all occupation numbers in terms of the qKdV charges, and confirmed that the qKdV charges can lift all degeneracies in a fixed conformal family but cannot distinguish $\Delta$.

The other parts in this paper concerned the central objective of testing GETH. We reviewed the basic principles of generalized eigenstate thermalization, reduced the computation of correlation functions to a free-bosonic field problem, obtained the first-order perturbations, attempted to express the expectation values of quasi-primaries as functions of qKdV charges, derived the reconstruction formula of occupation numbers from qKdV charges, analyzed the distinguishability problem of microscopic qKdV simultaneous eigenstates, and mentioned that qKdV charges are complete in the scope of vacuum-family conserved charges. 

The infinite-tower GETH claims equivalence between a thermal state and a qKdV eigenstate, while the truncated GETH claims the absence of the need to know the full information of the eigenstate in this process. We found that the original version of truncated GETH fails at the subleading order, by finding explicit counterexamples, and we argued for the validity of the general version of GETH considering all qKdV charges, by demonstrating that the difference between two admissible values of $\tilde{\Delta}$ must be $\ell$-subleading at fixed qKdV charges. The failure of truncated GETH matches exactly with the prediction in Ref.~\cite{Novaes:2021vjh}. Actually, Ref.~\cite{Novaes:2021vjh} claimed that GETH is correct for the leading order, i.e.~true in the strict large-$c$ limit; but local observables at finite levels are not likely to be determined by a finite number of qKdV charges just up to the $1/c$ order. Possible prospect of modifications of truncated GETH were also proposed. 

Next, we discuss the dilemmas and prospects of GETH. To introduce the discussion, we review the distinguishability of microstates by qKdV charges: the complete set of qKdV charges does not provide complete microstate distinguishability across different conformal families, but does provide that in a specific conformal family. In conclusion, the qKdV charges in the GGE retain most of the information about the microstates. This raises a naturalness concern: an ensemble is supposed to encode microstates using only a small amount of information, whereas the GGE does the opposite. In other words, the thermal states represented by the GGE are so largely non-thermal that they are close to individual microstates. In fact, numerous studies have raised philosophical doubts over the GGE--GETH framework, where they question the paradigmatic status of GGEs for statistical description, their physical necessity, and their substantive distinction from Gibbs ensembles~\cite{Anza:2018xmq,Senese:2026ykd}. Moreover, ETH-based studies of 2D CFTs have continued to develop since the proposal of GETH, implying that equivalent descriptions in the languages of Gibbs ensemble--ETH and GGE--GETH might be established in some thermal properties of 2D CFTs~\cite{Collier:2019weq,Numasawa:2022cni,Geng:2025efs}. These developments temper the case for regarding GETH as a distinct paradigm for describing thermal phenomena in 2D CFTs to some degree. 

However, in the context of 2D large-$c$ GGE, the qKdV charges in the expressions of finite-level observables truncate at the leading order in $c$~\cite{Dymarsky:2019etq,Novaes:2021vjh,Chen:2024lji}. Consequently, for some specific problems, only the first few qKdV charges in the GGE play an effective role, which resolves the non-thermality dilemma. Therefore, an important question for future studies of GETH is whether the expression of observables in qKdV charges admits a universal truncation scheme at subleading orders in the large-$c$ expansion, compatible with that at the leading order, thereby avoiding the need to reconstruct the microscopic details of the state, as discussed in Section \ref{sec:outcomes}. This would require knowledge of the qKdV spectrum at higher orders in the $1/c$ expansion, and perhaps even an exact determination of it. 

A few other foundational problems of the qKdV--GGE--GETH framework remain unresolved. The first is the existence of the inverse map $\{q_{2k-1}\}\mapsto\{\mu_{2k-1}\}$~\cite{Downing:2023lnp}, which has been mentioned repeatedly throughout this paper, and we do not repeat its details again. Another prospect is about the off-diagonal elements: while GETH focuses on diagonal matrix elements of local observables, the off-diagonal elements, relevant to relaxation and temporal fluctuations, have also been studied in 2D CFTs in the context of ETH and GETH~\cite{Brehm:2018ipf,Besken:2019bsu} and in another integrable system~\cite{Orlov:2026oox}, providing useful formulations for future studies of GGE-related thermal behavior of local-operator off-diagonal matrix elements in 2D CFTs. Finally, the extension of GETH to non-vacuum-family observables and conserved charges remains largely unexplored~\cite{Novaes:2021vjh,Downing:2021mfw}.

To conclude, we highlight the main contributions of this work beyond previous studies of GETH. It provides a comprehensive analysis of the general spectral structures of qKdV charges and vacuum-family observables, develops a framework for studying observable expectation values at subleading orders in the large-$c$ expansion for the first time, and provides the first quantitative analysis of the infinite tower of qKdV charges as a whole, all of which serve as preliminaries of discussions on truncated GETH and infinite-tower GETH. In addition, the problems to be addressed in this work lead to a proof of the generic non-degeneracy of qKdV charges in any specific conformal family, as well as a quantitative analysis of the extent to which qKdV charges can distinguish microstates. Our results may also offer several useful directions for future studies of GETH. First of all, our formulation based on the free-bosonic Fock space may be applied to future research on diagonal and off-diagonal matrix elements of observables in 2D CFTs and other integrable systems at higher orders. The structure of observable expectation values at subleading orders in the large-$c$ expansion, derived in this work, may provide a starting point for further investigations of qKdV charge truncations as well. We also hope that our treatment of the infinite tower of qKdV charges will offer valuable insights into the inverse problem of reconstructing chemical potentials from qKdV charges. 

\section*{Acknowledgments}

JZ acknowledges support from NSFC Grant No.~12205217 and Tianjin University Self-Innovation Fund Extreme Basic Research Project (Grant No.~2025XJ21-0007).

ChatGPT (5.6, 6, and 6.1) was used in this work. Mathematical derivations in Appendix \ref{app:nondeg} were assisted by it, while those in Appendix \ref{app:sol} were partially AI-assisted, with the central ideas and key steps guiding the derivation developed independently by the authors. It also served as a indispensable discussion partner in Section \ref{sec:scalings}. In Sections \ref{sec:cor} and \ref{sec:expectations}, it was used only for mechanical calculations based on the authors' formulations. All physical ideas are contributed by the authors, and all AI-assisted outputs have been checked by them.

\appendix

\section{PBW states as the leading orders of qKdV eigenstates}\label{app:pbw}

We prove that PBW states are exactly the leading orders of qKdV eigenstates in this appendix. The term ``PBW descendants/states'' refers to the following descendant states
\begin{equation}
|\Delta,\,\{n_k\}\rangle\rangle_{\mathrm{PBW}}=\left(L_{-k_{max}}\right)^{n_{k_{max}}}\cdots\left(L_{-2}\right)^{n_2}\left(L_{-1}\right)^{n_1}|\Delta\rangle.
\end{equation}
Only $Q_1,\,Q_3,\,c,\,\Delta,\,|\Delta\rangle,\,D$ follow the notations introduced in the main text here.

We now consider the operator $D=\sum_{l=1}^\infty L_{-l}L_l$. The simultaneous eigenstates of $Q_1$ and $D$ are precisely those of $Q_1$ and $Q_3$, which are in fact simultaneous eigenstates of all qKdV charges. At a fixed level $N$, a basis of the linear subspace of $Q_1$ eigenstates can be formally arranged in PBW order as the row vector
\begin{equation}
\mathbf{V}=\{V_i\}=
( L_{-N}|\Delta\rangle,\,L_{-(N-1)}L_{-1}|\Delta\rangle,\,\cdots,\,L_{-2}L_{-1}^{N-1}|\Delta\rangle,\,L_{-1}^N|\Delta\rangle ),
\end{equation}
with the corresponding coefficients collected into the column vector $\mathbf{\Lambda}=\{\lambda_i\}$. Here, the PBW order is the convention of arranging all PBW states in the same level-$N$ linear subspace from left to right according to the descending order of the indices of the Virasoro generators appearing in the states, starting from the leftmost one, then the next one to the left, and so on. The PBW order is widely used because it leads to clear results~\cite{Kac:2013bl}. By linearity of $D$, let $\mathbf{D}_N$ denote the matrix representing the action of $D$ on this basis, such that
\begin{equation}\label{eqt:opemat}
D\mathbf{V}=\mathbf{V}\mathbf{D}_N .
\end{equation}
And then the eigenvalue equation
\begin{equation}
D\left(\mathbf{V}\mathbf{\Lambda}\right)\propto\mathbf{V}\mathbf{\Lambda}
\end{equation}
can be reduced. The left-hand side becomes $D\left(\mathbf{V}\mathbf{\Lambda}\right)=\sum_i\lambda_iDV_i=\mathbf{V}\mathbf{D}_N\mathbf{\Lambda}$, and therefore, $\mathbf{V}\mathbf{D}_N\mathbf{\Lambda}\propto\mathbf{V}\mathbf{\Lambda}$, which gives
\begin{equation}
\mathbf{D}_N\mathbf{\Lambda}=\lambda\mathbf{\Lambda}.
\end{equation}

We next examine the action of $D$ on a PBW state:
\[
\sum_{l=1}^\infty L_{-l}L_l\ L_{-k_{max}}^{n_{k_{max}}} \cdots L_{-1}^{n_1}|\Delta\rangle.
\]
It maps a given level-$N$ PBW state just into the same linear subspace. We focus on the action of $L_l$ on a PBW state. It is easy to see that it gives a nonzero contribution only when $l\leq N$, and hence the summation can be truncated at $l=N$. Starting from $L_l$, the Virasoro algebra can be used to move the Virasoro generators with nonnegative indices to the right, thereby generating new terms. The same procedure is then applied to each new term. Together with the condition $L_0|\Delta\rangle=\Delta|\Delta\rangle$, this process is iterated until no Virasoro generator with a nonnegative index remains in the expression.

Since the commutation of $L_k$ and its subsequent counterparts can generate at most one $L_0$ and one central term, while the commutator of $L_0$ with Virasoro generators of negative indices only contributes terms independent of $c$ and $\Delta$, $\mathbf{D}_N$ is at most linear in $(c,\,\Delta)$, namely,
\begin{equation}
\mathbf{D}_N=\mathbf{K}_N+\Delta\mathbf{H}_N+c\mathbf{C}_N .
\end{equation}
Recall that the discussion is in the large $(c,\,\Delta)$ limit. Therefore, at leading order, any terms independent of $c$ and $\Delta$ contributing to the entries of $\mathbf{D}_N$ can be neglected, i.e.~$\mathbf{D}_N\simeq\Delta\mathbf{H}_N+c\mathbf{C}_N$.

Since only Virasoro commutators of the form $[L_m,\,L_{-m}]$ contribute to $\Delta$ or $c$ terms, the total contributions of $L_l$ to $\mathbf{H}_N$ and $\mathbf{C}_N$ can be obtained as weighted sums over the integer partitions of $l$. It is known that $\sum_{k=1}^\infty k n_k=N$; and let $\{s_k\}$ denote the number of times that $k$ appears in a partition of $l$, so that $\sum_{k=1}^\infty k s_k=l$, $s_k\leq n_k$. We further introduce an auxiliary sequence of positive integers $\{\alpha_j\}$, whose physical information is equivalent to $\{s_i\}$, with its largest index denoted by $m$, by listing the $k$'s in descending order according to their multiplicities. Finally, we introduce another auxiliary sequence of positive integers $\{a_j\}$ satisfying $a_1=k,\,a_j=a_{j-1}-\alpha_{j-1}$. Define
\begin{numcases}{}
Q(k,\,\alpha)=\prod_{j=1}^m(a_j+\alpha_j),\\
R(k,\,\alpha)=\frac{a_m^3-a_m}{12}\prod_{j=1}^{m-1}(a_j+\alpha_j).
\end{numcases}
Denote $|\Delta,\,\{n_k\}\rangle_{\mathrm{PBW}}\mathrel{:=}L_{-k_{max}}^{n_{k_{max}}} \cdots L_{-1}^{n_1}|\Delta\rangle$ as before, and then one obtains
\begin{numcases}{}
H|\Delta,\,\{n_k\}\rangle_{\mathrm{PBW}}=\sum_{k=1}^N\left[\sum_\alpha Q(k,\,\alpha)\left(\prod_{i=1}^N \binom{n_i}{s_i}\right)L_{-k}L_{-k_{max}}^{n_{k_{max}}-s_{k_{max}}} \cdots L_{-1}^{n_1-s_1}|\Delta\rangle\right],\label{eqt:h}\\
C|\Delta,\,\{n_k\}\rangle_{\mathrm{PBW}}=\sum_{k=1}^N\left[\sum_\alpha R(k,\,\alpha)\left(\prod_{i=1}^N \binom{n_i}{s_i}\right)L_{-k}L_{-k_{max}}^{n_{k_{max}}-s_{k_{max}}} \cdots L_{-1}^{n_1-s_1}|\Delta\rangle\right],\label{eqt:c}
\end{numcases}
in which $H$ and $C$ are the operators corresponding to $\mathbf{H}_N$ and $\mathbf{C}_N$.

It is observed that all the terms in Eqs.~(\ref{eqt:h}) and (\ref{eqt:c}) either preserve $\{n_k\}$ or join some of the occupied $k$'s into a single larger $k$. During the process of moving the Virasoro generators with nonnegative indices to the right, apart from the terms obtained by directly moving $L_{-l}$ to the right, all other terms involve similar joinings. Therefore, all new terms generated by the action of $D$ make the partition of $N=\sum_{k=1}^Nkn_k$ coarser. Corresponding to Eq.~(\ref{eqt:opemat}), the leading order of $\mathbf{D}_N$ is an upper triangular matrix.

The eigenvalues of an upper triangular matrix are given by its diagonal entries. Consider
\[
L_{-l}L_l L_{-k_{max}}^{n_{k_{max}}-s_{k_{max}}} \cdots L_{-1}^{n_1-s_1}|\Delta\rangle .
\]
Only when $l$ equals the absolute value of the index of one of the generators does the term contribute to the diagonal entry. In this case, $Q(k,\,\alpha)=2k$, $R(k,\,\alpha)=\frac{k^3-k}{12}$, and $\prod_{i=1}^N \binom{n_i}{s_i}=n_k$. For the column $a$ corresponding to this state, one has
\begin{equation}\label{eqt:hc}
\begin{cases}
\displaystyle\left(\mathbf{H}_N\right)_{aa}=2N,\\
\displaystyle\left(\mathbf{C}_N\right)_{aa}=\sum_{k=1}^{N}n_k\frac{k^3-k}{12},
\end{cases}
\end{equation}
which aligns with Ref.~\cite[Section 2.2.1]{Dymarsky:2018lhf}. This leads to accidental degeneracies in the eigenvalues of $Q_3$, e.g.~$L_{-3}L_{-1}^7|\Delta\rangle$ and $L_{-2}^4L_{-1}^2|\Delta\rangle$ for the case of finite level, which is also the fundamental source of the complexity involved in the discussion of Section \ref{sec:spectrum}.

Solving the eigenvalue equation of an upper triangular matrix is equivalent to successively solving the eigenvalue equations of the upper-left $i\times i$ submatrices, with all $\lambda_i$'s below the corresponding position filled with zero when the eigenvector is restored to its original dimension. On the other hand, different basis states with the same number of Virasoro generators are not connected by $\mathbf{D}_N$, and hence reversing their ordering still leaves $\mathbf{D}_N$ triangular.

Finally, we observe that the leading-order contribution to the inner product of PBW states is
$$
\prod_{k=1}^{N}n_k!\left([L_k,\,L_{-k}]\right)^{n_k}\sim O\left(c^{\sum_{k=1}^Nn_k}\right).
$$

Thus, for each eigenvector, the contributions from basis states other than that in the $a$-th column become either zero or negligible. Therefore, the simultaneous eigenstates of $Q_1$ and $Q_3$ are PBW states at leading order, which can be normalized as
\begin{equation}
|\Delta,\,\{n_k\}\rangle_{\mathrm{PBW}}=\prod_{k=1}^{N}\frac{1}{\sqrt{n_k!}}\left(\frac{L_{-k}}{\sqrt{[L_k,\,L_{-k}]}}\right)^{n_k}|\Delta\rangle+\cdots,
\end{equation}
and their corresponding $D$-eigenvalues are directly given by the sum of $\Delta\left(\mathbf{H}_N\right)_{aa}+c\left(\mathbf{C}_N\right)_{aa}$ corresponding to this state. The $Q_3$ eigenvalues are
\begin{equation}
Q_3=\Delta^2+6N\Delta+N^2+\frac{c}{6}\sum_{k=1}^{N} n_k(k^3-k)-\left(\frac{c}{12}+\frac{1}{6}\right)(\Delta+N)+\frac{c}{24}\left(\frac{c}{24}+\frac{1}{60}\right)+\cdots.
\end{equation}

The applicability of this approximation requires $N$ to be sufficiently small such that the terms independent of $c,\,\Delta$ are not large enough to alter the leading-order behavior determined by the terms linear in $(c,\,\Delta)$. Since the physical scaling~\cite{Datta:2019jeo,Dymarsky:2019etq} involves the large-$c$ limit and $\Delta\sim c\ell^2$ together with the absence of $c$ in the scaling of occupation-number moments, the PBW states can indeed be treated as the leading order with respect to $c$. This is also why this line of reasoning is followed in the discussion of Section \ref{sec:freeboson}, replaced by the application of Fock states. This also explains why PBW states look inconsistent with the qKdV eigenstates in the scaling $M_{2k-1}\sim c\ell^{2k}$~\cite{Guo:2018pvi}.

\section{On the measure zero of degeneracy curves of first two qKdV charges}\label{app:nondeg}
In this appendix, the measure-zero nature of the region in the $(c,\,\Delta)$ parameter space where the first two qKdV charges are degenerate is proved.

To avoid confusion, the $\lambda$ in Section \ref{sec:note} is relabelled as $q$ in this appendix. For a polynomial $P(q)=a_n\prod_{i=1}^{n}(q-q_i)$, the discriminant of its roots is
\begin{equation}
\operatorname{disc}(P)=a_n^{2n-2}\prod_{1\leq i<j\leq n}(q_i-q_j)^2 .
\end{equation}
We need to show that for all levels $N$, the discriminant of the roots of $\det(\mathbf{D}_N-q\mathbf{I})=0$ is not the zero polynomial in $(c,\,\Delta)$.

Consider the leading order of physical quantities in the sense of $(c,\,\Delta)$. For $N\leq 7$, the simultaneous eigenstates of $Q_1$ and $Q_3$ are non-degenerate; for $N\geq 8$, degeneracies are inevitable, because the two structures $L_{-3}L_{-1}^5$ and $L_{-2}^4$ contribute to the same $Q_3$. We will next consider exact results for the discriminant of the eigenvalues beyond the large-$(c,\,\Delta)$ limit, in order to analyze whether this degeneracy is merely due to information lost in the approximation or is genuinely structural. The result is mainly related to the former and less related to the latter, since the degeneracy region consists of countably infinite one-dimensional curves, as Theorem \ref{theo:1} claimed. 

We introduce the notations in this appendix in advance. We first review those defined in Section \ref{sec:note}: 
\[
Q_3 = 2D+L_0^2-\frac{c+2}{12}L_0 +\frac{c(5c+22)}{2880}I, \quad D:=\sum_{n\ge1}L_{-n}L_n.
\]
For each $N\ge0$ in any conformal family, we let $\mathcal{F}_N$ denote the level-$N$ subspace, and define
\[
D_N(c,\,\Delta) := D\big|_{\mathcal{F}_N}.
\]
It is obvious that the matrices $\mathbf{D}_N$ are just those of the operators $D_N$. The discussed determinant is defined as
\[
\operatorname{disc}_N(c,\,\Delta) := \operatorname{Disc}_q\chi_N(q;\,c,\,\Delta),\quad\chi_N(q;\,c,\,\Delta):=\det\left(\mathbf{D}_N-q\mathbf{I}\right).
\]
Moreover, we retain the notation Eqs.~(\ref{eqt:not}): $\tilde c=\frac{c-1}{24}, \ \tilde\Delta=\Delta-\tilde c$, and we define
\[
P:=\sqrt{2\tilde\Delta}
\]
additionally for convenience. 

As Section \ref{sec:note} mentioned, it is required that $\operatorname{disc}_N$ is not the zero polynomial, which reduces to proving
\begin{equation}\label{eqt:simple}
\operatorname{disc}_N(c,\,\Delta)\not\equiv0, \ \forall\,N\ge0,
\end{equation}
and further to finding a non-degenerate point at each $N$. The purpose of this appendix is to establish this property at every level $N$. The Virasoro commutation relations and the PBW basis imply that the coefficients of $\chi_N$ lie in $\mathbb Q[c,\,\Delta]$, and hence $\operatorname{disc}_N\in\mathbb Q[c,\,\Delta]$. We therefore construct the following explicit family of parameter values at which $D_N$ has simple spectrum. 
\begin{lemma}\label{lem:3}
For every integer $N\ge0$ and every real transcendental number $P_*$, there exists a real number $\varepsilon_N(P_*)>0$ such that, for every real $s$ satisfying $0<|s|<\varepsilon_N(P_*)$, the operator $D_N$ has simple spectrum at $c=-2+6s,\ \Delta=\frac{P_*^2}{2}-\frac18+\frac s4$.
\end{lemma}

In particular, $\operatorname{disc}_N(c,\,\Delta)\neq0$ at these parameter values. Thus, for each fixed level $N$, the degeneracy locus
\[
\left\{ (c,\,\Delta)\in\mathbb R^2: \operatorname{disc}_N(c,\,\Delta)=0 \right\}
\]
is a proper algebraic subset. The parameters exhibiting degeneracy at some level therefore lie in a countable union of proper algebraic subsets, which has Lebesgue measure zero in the real parameter plane.
\subsection{Review on $c=-2$ spectrum}
For $k>0$, denote the quadratic oscillator terms in the Virasoro generators of Eq.~(\ref{eqt:rea}) of the main text by
\[
B_k := \frac{1}{2}\sum_{p=1}^{k-1}a_p a_{k-p}+\sum_{p=1}^{\infty}a_{-p}a_{k+p}, \quad B_{-k} := \frac{1}{2}\sum_{p=1}^{k-1}a_{-p} a_{-(k-p)}+\sum_{p=1}^{\infty}a_{-(k+p)}a_{p}.
\]
Then
\begin{equation}\label{eqt:free-field}
L_{\pm k}=\left(P\pm i k\sqrt{2\tilde c}\right)a_{\pm k}+B_{\pm k},\quad L_0=\Delta+\mathcal N,
\end{equation}
where
\[
\mathcal N:=\sum_{k=1}^{\infty}a_{-k}a_k, \quad \left.\mathcal N\right|_{\mathcal{F}_N}=NI.
\]

In the free-bosonic realization, any set of bosonic occupation numbers of the Fock states is a partition of the level $N$, which $\lambda$ and $\mu$ henceforth both denote, i.e.~$\lambda=\{\lambda_i\},\ \mu=\{\mu_i\} \vdash N$. As preparation, the shifted power sums are defined by
\[
\mathsf p_k(\lambda) := \sum_{i\ge1} \left[ \left( \lambda_i-i+\frac12 \right)^{k-1} - \left( -i+\frac12 \right)^{k-1} \right], \ k\ge2.
\]
This sum has only finitely many nonzero terms, and $\mathsf p_2(\lambda)=N$. We first set $\tilde c=-\frac18$, so that $c=-2$. The diagonalization result~\cite{Masoero:2026rvz} states that the Schur functions $s_\lambda$, $\lambda\vdash N$, form an eigenbasis of $Q_3$ on $\mathcal{F}_N$, with
\[
Q_3s_\lambda=\left[\left(\frac{P^2}{2}-\frac18\right)^2+3P^2N+3P\mathsf p_3(\lambda)+\mathsf p_4(\lambda)-\frac1{120}\right]s_\lambda,
\]
which equals
\begin{equation}\label{eqt:D_N formula}
D_Ns_\lambda=\left(P^2N+\frac N8-\frac{N^2}{2}+\frac32P\mathsf p_3(\lambda)+\frac12\mathsf p_4(\lambda)\right)s_\lambda.
\end{equation}
And at this parameter, $\Delta=\frac{P^2}{2}-\frac18$. At fixed level $N$, after subtracting a common scalar independent of $\lambda$, the quantity
\[
E_\lambda(P) := \mathsf p_4(\lambda) + 3P\mathsf p_3(\lambda)
\]
completely determines the unperturbed spectrum. 

Varying $P$ alone cannot remove all degeneracies. If distinct partitions $\lambda,\,\mu\vdash N$ satisfy $\mathsf p_3(\lambda)=\mathsf p_3(\mu), \ \mathsf p_4(\lambda)=\mathsf p_4(\mu)$, then $E_\lambda(P)\equiv E_\mu(P)$ as functions of $P$. For example, at level~15, the partitions
\begin{equation}\label{eqt:counter}
\lambda=(5,\,2,\,2,\,2,\,2,\,2),\ \mu=(4,\,4,\,2,\,2,\,1,\,1,\,1),
\end{equation}
satisfy $\mathsf p_3(\lambda) = \mathsf p_3(\mu) = -30$, $\mathsf p_4(\lambda) = \mathsf p_4(\mu) = \frac{1395}{4}$. We therefore fix $P$ and examine the dependence of $D_N$ on $\tilde c$. Define
\[
M := \sum_{k=1}^{\infty} k^2a_{-k}a_k.
\]
Eqs.~(\ref{eqt:free-field}) gives
\[
D_N =\sum_{k=1}^{\infty} B_{-k}B_k + P\sum_{k=1}^{\infty} \left( B_{-k}a_k+a_{-k}B_k \right)+ i\sqrt{2\tilde c} \sum_{k=1}^{\infty} k\left( B_{-k}a_k-a_{-k}B_k \right) + P^2NI + 2\tilde c\,M . 
\]
In the identity $\sum_{k=1}^{\infty} k\bigl( B_{-k}a_k-a_{-k}B_k \bigr) = \frac12 \sum_{i,\,j=1}^{\infty} (j-i) \bigl( a_{-i}a_{-j}a_{i+j} - a_{-(i+j)}a_ia_j \bigr)$, the operator $\bigl( a_{-i}a_{-j}a_{i+j} - a_{-(i+j)}a_ia_j \bigr)$ is symmetric in $i,\,j$, whereas the coefficient $j-i$ is antisymmetric. Hence, $\sum_{k=1}^{\infty} k\bigl( B_{-k}a_k-a_{-k}B_k \bigr) = 0$. We obtain the decomposition
\begin{equation}\label{eqt:osc}
D_N=P^2NI+P\sum_{k=1}^{\infty}\bigl(B_{-k}a_k+a_{-k}B_k\bigr)+2\tilde cM+\sum_{k=1}^{\infty}B_{-k}B_k.
\end{equation}
Thus, at fixed $P$, the dependence of $D_N$ on $\tilde c$ is entirely contained in $2\tilde cM$.

Now set $\tilde c = -\frac18+\frac s4$. Combining Eqs.~(\ref{eqt:D_N formula}) and (\ref{eqt:osc}), we can write $D_N$ as
\[
D_N = \left( P^2N+\frac N8-\frac{N^2}{2} \right)I + \frac12\widehat D_N(P,\,s),
\]
where
\begin{equation}\label{eqt:reduced-D}
\widehat D_N(P,\,s):=\widehat D_N(P,\,0)+sM,\quad\widehat D_N(P,\,0)s_\lambda:=E_\lambda(P)s_\lambda.
\end{equation}
The corresponding parameters are $c =-2+6s$, $\Delta = \frac{P^2}{2}-\frac{1}{8}+\frac{c}{4}$. Obviously, the operators $D_N$ and $\widehat D_N(P,\,s)$ have the same spectral multiplicities for fixed $(P,\,N)$, reducing the simple-spectrum problem to the study of $\widehat D_N(P,\,s)$.

Finally, we describe the degeneracies of $\widehat D_N(P,\,0)$. For $\lambda,\,\mu\vdash N$, define
\[
\lambda\sim\mu \quad\Longleftrightarrow\quad \bigl( \mathsf p_3(\lambda),\,\mathsf p_4(\lambda) \bigr) = \bigl( \mathsf p_3(\mu),\,\mathsf p_4(\mu) \bigr).
\]
Equivalently, $E_\lambda(P)\equiv E_\mu(P)$ holds as a polynomial identity in $P$. Denote an equivalence class by $\mathcal G$, and let $\mathcal F_{\mathcal G} := \operatorname{span}_{\mathbb C} \{s_\lambda:\lambda\in\mathcal G\}$ be the span of the corresponding Schur functions. For $k=3,\,4$, denote $\mathsf p_k(\mathcal G) := \mathsf p_k(\lambda),\,\lambda\in\mathcal G$, and define $E_{\mathcal G}(P) := \mathsf p_4(\mathcal G) + 3P\mathsf p_3(\mathcal G)$. 

If $\mathcal G\neq\mathcal G'$, then $E_{\mathcal G}$ and $E_{\mathcal G'}$ are distinct affine functions with rational coefficients, so they can coincide at no more than one rational point. Define
\[
\Sigma_N := \left\{ P\in\mathbb R\,:\,\exists\, \mathcal G\neq\mathcal G', \  E_{\mathcal G}(P) = E_{\mathcal G'}(P) \right\}.
\]
At fixed $N$, there are finitely many equivalence classes, so $\Sigma_N\subset\mathbb Q$ is finite. Throughout the following discussion, we assume $P\notin\Sigma_N$. Then $\mathcal F_{\mathcal G}$ is precisely the full eigenspace of $\widehat D_N(P,\,0)$ associated with $E_{\mathcal G}(P)$. It remains to determine whether the perturbation $sM$ removes the degeneracy within each $\mathcal F_{\mathcal G}$.
\subsection{Maya diagram formulation}\label{app:matrix-elements}
To describe the matrix elements of $M$ in the Schur basis, we use the standard symmetric-function realization of the Heisenberg algebra:
\[
a_{-k} \mapsto x_k, \ a_k \mapsto k\frac{\partial}{\partial x_k}, \ k>0.
\]
Hereafter, $s_\lambda$ denotes the Schur basis in this realization. Equip the space with the Hermitian inner product for which the Schur basis is orthonormal, $\langle s_\lambda,\,s_\mu\rangle=\delta_{\lambda\mu}$, with the convention that the first argument is conjugate-linear. Then $a_k^\dagger=a_{-k}$, $M^\dagger=M$. In particular, $\widehat D_N(P,\,s)$ is self-adjoint for real $(P,\,s)$. We next introduce Maya diagrams.
\begin{definition}
Let $\mathbb Z+\frac12$ denote the set of half-integers. For a partition $\lambda=(\lambda_1\ge\lambda_2\ge\cdots)$, the Maya diagram of $\lambda$ is the pair $(\operatorname{Occ}(\lambda),\,\operatorname{Vac}(\lambda))$, where
\[
\operatorname{Occ}(\lambda)=\left\{\lambda_i-i+\frac12:\ i\ge1\right\}\subset\mathbb Z+\frac12,  \quad\operatorname{Vac}(\lambda)=\mathbb Z+\frac12\setminus\operatorname{Occ}(\lambda).
\]
These are called the occupied set and the vacancy set, respectively.
\end{definition}

The occupied set of the Maya diagram of the empty partition $\varnothing$ is
\[
\operatorname{Occ}(\varnothing)=\left\{-\frac12,\,-\frac32,\,-\frac52,\,\cdots\right\}.
\]
The Maya diagram of any partition $\lambda$ differs from the vacuum Maya diagram at only finitely many sites, so the symmetric difference $\operatorname{Occ}(\lambda)\triangle\operatorname{Occ}(\varnothing)$ is finite. Thus, $\mathsf p_k(\lambda)$ can be written as
\[
\mathsf p_k(\lambda)=\sum_{x\in\operatorname{Occ}(\lambda)\setminus\operatorname{Occ}(\varnothing)}x^{k-1}-\sum_{x\in\operatorname{Occ}(\varnothing)\setminus\operatorname{Occ}(\lambda)} x^{k-1},\ k\ge2.
\]

We now describe transitions between Maya diagrams. Let $\lambda,\,\mu$ be partitions. Suppose there exist $\alpha>\beta$, $\alpha,\,\beta\in\operatorname{Occ}(\lambda)$, and $\gamma>\delta$, $\gamma,\,\delta\in\operatorname{Vac}(\lambda)$, such that $\operatorname{Occ}(\mu)=\bigl(\operatorname{Occ}(\lambda)\setminus\{\alpha,\,\beta\}\bigr)\cup\{\gamma,\,\delta\}$, then we say that a transition from $\lambda$ to $\mu$ occurs, and write
\[
\{\alpha,\,\beta\}\longrightarrow\{\gamma,\,\delta\}.
\]
Thus, the two particles at $(\alpha,\,\beta)$ move to the vacant sites $(\gamma,\,\delta)$.

\begin{figure}[htbp]
\centering
\[
\begin{array}{c|rrrrrrrr}

\textrm{Partition}&-\frac72&-\frac52&-\frac32&-\frac12&\phantom{-}\frac12&\phantom{-}\frac32&\phantom{-}\frac52&\phantom{-}\frac72\\ \hline

\varnothing&\bullet&\bullet&\bullet&\bullet&\circ&\circ&\circ&\circ\\
(3,1)&\bullet&\bullet&\circ&\bullet&\circ&\circ&\bullet&\circ
\end{array}
\]
\caption{Some examples of Maya diagrams.}
\end{figure}

The diagonal and non-diagonal formulae below are obtained from Ref.~\cite{vanIttersum:2024mqe} by conversion to our normalization.
\begin{lemma}
In our normalization, the diagonal entries of $M$ in the Schur basis are
\begin{equation}\label{eqt:diagonal-M}
M_{\lambda\lambda}=N^2-\frac N{12}+\frac{\mathsf p_4(\lambda)}3.
\end{equation}
For the off-diagonal entries $M_{\mu\lambda},\,\forall \lambda\ne\mu $, the only possible transitions are
\[
\{\alpha,\,\beta\}\longrightarrow\{\gamma,\,\delta\}, \ \alpha>\beta, \ \gamma>\delta, \ \alpha+\beta=\gamma+\delta,
\]
and if $\alpha,\,\beta$ have indices $i<j$ in the decreasingly ordered initial occupied set, and $\gamma,\,\delta$ have indices $r<t$ in the decreasingly ordered final occupied set, then
\begin{equation}\label{eqt:signed-M}
M_{\mu\lambda}=(-1)^{i+j+r+t}\bigl((\alpha-\gamma)^2-(\alpha-\delta)^2\bigr),
\end{equation}
and hence
\begin{equation}\label{eqt:amplitude}
|M_{\mu\lambda}|^2=(\alpha-\beta)^2(\gamma-\delta)^2.
\end{equation}
\end{lemma}

To determine the effect of such a transition on $\mathsf p_i$ and $E_\lambda(P)$, denote
\begin{equation}\label{eqt:S}
S:=\alpha+\beta=\gamma+\delta, \quad x:=\alpha-\beta>0, \quad y:=\gamma-\delta>0.
\end{equation}
Since the transition replaces the occupied sites $(\alpha,\,\beta)$ by $(\gamma,\,\delta)$, the definition of the shifted power sums gives
\begin{equation}\label{eqt:gap}
\begin{cases}
\displaystyle\mathsf p_3(\lambda)-\mathsf p_3(\mu)=\alpha^2+\beta^2-\gamma^2-\delta^2=\frac{x^2-y^2}{2}, \\
\displaystyle\mathsf p_4(\lambda)-\mathsf p_4(\mu)=\alpha^3+\beta^3-\gamma^3-\delta^3=\frac{3S(x^2-y^2)}4.
\end{cases}
\end{equation}
Therefore,
\[
E_\lambda(P)-E_\mu(P)=\frac32(x^2-y^2)\left(P+\frac S2\right).
\]
A nontrivial transition necessarily satisfies $x\ne y$. Otherwise, $x=y$ and $\alpha+\beta=\gamma+\delta$ would imply $\{\alpha,\,\beta\}=\{\gamma,\,\delta\}$, contradicting nontriviality. Thus, if $M_{\mu\lambda}\ne 0$, then $ \mathsf p_3(\lambda)\ne \mathsf p_3(\mu)$. In particular, for $\lambda,\,\mu \in \mathcal G$,
\[
\lambda\ne\mu \quad\Longrightarrow\quad M_{\mu\lambda}=0.
\]
On the other hand, Eq.~(\ref{eqt:diagonal-M}) shows that, for every equivalence class $\mathcal{G}$ and every $\lambda \in \mathcal{G}$, $M_{\lambda\lambda}$ equals the same constant 
\begin{equation}\label{eqt:copy}
u_{\mathcal G} = N^2-\frac N{12} +\frac{\mathsf p_4(\mathcal G)}3.
\end{equation}
\subsection{The second-order effective operator and its properties}\label{app:first-order}
\subsubsection{Introduction}
We study the first- and second-order perturbations of $\widehat D_N(P,\,s)=\widehat D_N(P,\,0)+sM$ on the full eigenspace $\mathcal F_{\mathcal G}$ and derive the second-order effective operator $K$.

Fix $P\notin\Sigma_N$ and an equivalence class $\mathcal G$, and denote $E^{(0)}:=E_{\mathcal G}(P)$. Then $\mathcal F_{\mathcal G}$ is the full eigenspace of $\widehat D_N(P,\,0)$ associated with $E^{(0)}$. By finite-dimensional analytic perturbation theory~\cite[Page 115--116]{Kato:1995pt}, the eigenvalues issuing from $E^{(0)}$ admit analytic branches near $s=0$. Choose one such branch and denote
\begin{equation}\label{eqt:perturbation-expansion}
E(s)=E^{(0)}+sE^{(1)}+s^2E^{(2)}+O(s^3),
\end{equation}
where $E^{(1)}$ and $E^{(2)}$ are the first- and second-order corrections, respectively. Let $\Pi_{\mathcal G}$ be the orthogonal projection from $\mathcal F_N$ onto $\mathcal F_{\mathcal G}$, and set $\Pi_{\mathcal G}^{\perp}:=I-\Pi_{\mathcal G}$. With respect to the orthogonal decomposition $\mathcal F_N=\mathcal F_{\mathcal G}\oplus \mathcal F_{\mathcal G}^{\perp}$, denote
\[
\widehat D_{N,\,\perp}:=\widehat D_N(P,\,0)|_{\mathcal F_{\mathcal G}^{\perp}}, 
\]
and define
\[
U:=\Pi_{\mathcal G}M|_{\mathcal F_{\mathcal G}},\quad V:=\Pi_{\mathcal G}M|_{\mathcal F_{\mathcal G}^{\perp}},\quad W:=\Pi_{\mathcal G}^{\perp}M|_{\mathcal F_{\mathcal G}^{\perp}}.
\]
The block matrices are
\[
\widehat D_N(P,\,0) = \begin{pmatrix} E^{(0)}I&0\\ 0&\widehat D_{N,\,\perp} \end{pmatrix}, \quad M= \begin{pmatrix} U&V\\ V^\dagger&W \end{pmatrix}.
\]
As $P\notin\Sigma_N$, we have $E^{(0)}\notin\operatorname{Spec}(\widehat D_{N,\,\perp})$, where ``$\operatorname{Spec}()$'' denotes the set of spectra, so $E^{(0)}I-\widehat D_{N,\,\perp}$ is invertible on $\mathcal F_{\mathcal G}^{\perp}$. Let $x_0\in \mathcal F_{\mathcal G}$ be the eigenvector at $s=0$ of the chosen branch. The first-order degenerate perturbation equation is $Ux_0=E^{(1)}x_0$. Eq.~(\ref{eqt:copy}) gives that $E^{(1)}=u_{\mathcal G}$ is common to all eigenvalue branches issuing from $E^{(0)}$. First-order perturbation therefore does not remove the degeneracy within $\mathcal F_{\mathcal G}$.

Since the first-order effective operator $U=u_{\mathcal G}I$ is scalar, the degeneracy can first split at second order. Standard degenerate perturbation theory gives the second-order effective operator on $\mathcal F_{\mathcal G}$:
\begin{equation}\label{eqt:effective-K}
K:=V\bigl(E^{(0)}I-\widehat D_{N,\,\perp}\bigr)^{-1}V^\dagger\in\operatorname{End}(\mathcal F_{\mathcal G}).
\end{equation}
Here, $\operatorname{End}(\mathcal F_{\mathcal G})$ denotes the space of all linear maps from $\mathcal F_{\mathcal G}$ to itself. Indeed, for a zeroth-order eigenvector $x_0\in \mathcal F_{\mathcal G}$, the first-order component of the eigenvector in $\mathcal F_{\mathcal G}^{\perp}$ is $y_1=(E^{(0)}I-\widehat D_{N,\,\perp})^{-1}V^\dagger x_0$, and the second-order equation reduces to $Kx_0=E^{(2)}x_0$. Hence, the second-order corrections $E^{(2)}$ are the eigenvalues of $K$. 

We now compute the matrix elements of $K$ in the Schur basis. Denote $M_{\lambda\mu}:=\langle s_\lambda,\,Ms_\mu\rangle$. Since $\widehat D_N(P,\,0)$ is diagonal in this basis, for $\mu\notin\mathcal G$ we have $\widehat D_{N,\,\perp}s_\mu=E_\mu(P)s_\mu$. Moreover, for $\nu\in\mathcal G$, 
\[
V^\dagger s_\nu = \sum_{\mu\notin\mathcal G}M_{\mu\nu}s_\mu, \quad Vs_\mu = \sum_{\lambda\in\mathcal G}M_{\lambda\mu}s_\lambda.
\]
Substituting into Eq.~(\ref{eqt:effective-K}), we obtain, for any $\lambda,\,\nu\in\mathcal G$, 
\begin{equation}\label{eqt:K-matrix}
K_{\lambda\nu}(P)=\sum_{\mu\notin\mathcal G}\frac{M_{\lambda\mu}M_{\mu\nu}}{E^{(0)}-E_\mu(P)}.
\end{equation}
\subsubsection{Diagonality}\label{app:second-order}
We now study the second-order matrix elements $K_{\lambda\nu}(P)$ in Eq.~(\ref{eqt:K-matrix}), where $\lambda,\,\nu\in\mathcal G$. 

\begin{lemma}\label{lem:second-diagonal}
If $\lambda\ne\nu$ and $\lambda, \,\nu\in\mathcal G$, then $K_{\lambda\nu}(P)\equiv0$.
\end{lemma}

\begin{proof}
Fix $\lambda\neq\nu\in\mathcal G$. By Eq.~(\ref{eqt:K-matrix}), each nonzero summand corresponds to a two-step path
\[
\lambda\xrightarrow{\tau_1}\mu\xrightarrow{\tau_2}\nu, \ \ \mu\notin\mathcal G .
\]
Its contribution to $K_{\lambda\nu}(P)$ is $\frac{M_{\lambda\mu}M_{\mu\nu}}{E^{(0)}-E_\mu(P)}$. We show that performing $\tau_2$ before $\tau_1$ gives another path whose contribution has the same numerator and the opposite denominator. The two contributions therefore cancel, yielding $K_{\lambda\nu}(P)\equiv0$.

Let $T_i$ be the set of four Maya sites involved in $\tau_i$. The sites whose occupations differ between the initial and final states are precisely $T_1\mathbin\triangle T_2$. If $T_1\cap T_2\ne\varnothing$, then $|T_1\mathbin\triangle T_2|=8-2|T_1\cap T_2|\le6$, with the vertical bars used to denote the cardinality of a set in this article. The numbers of removed and added occupied sites are equal, so each is at most three. Since $\lambda,\,\nu\in\mathcal G$, they have the same $\mathsf p_2,\,\mathsf p_3,\,\mathsf p_4$, and hence
\begin{equation}\label{eqt:three-moments}
\sum_{\xi\in\operatorname{Occ}(\lambda)\setminus\operatorname{Occ}(\nu)}\xi^j=\sum_{\zeta\in\operatorname{Occ}(\nu)\setminus\operatorname{Occ}(\lambda)}\zeta^j,\ j=1,\,2,\,3.
\end{equation}
Newton's identities~\cite[Proposition~2.6.3]{Savage:2022gtg} imply that the two sets of sites coincide. Since they are disjoint, both are empty, contradicting $\lambda\ne\nu$. Thus, $T_1\cap T_2=\varnothing$. The transitions are independent, so interchanging their order gives a valid path
\[
\lambda\xrightarrow{\tau_2}\mu' \xrightarrow{\tau_1}\nu.
\]
We call this the reordered path associated with $\lambda\xrightarrow{\tau_1}\mu\xrightarrow{\tau_2}\nu$. Since $\tau_2$ is nontrivial, the matrix-element formulae of the preceding subsection give $\mathsf p_3(\mu')\neq\mathsf p_3(\lambda)$, so $\mu'\notin\mathcal G$. Thus, the reordered path also corresponds to a summand in Eq.~(\ref{eqt:K-matrix}). 

We compare the two contributions. Since $T_1$ and $T_2$ are disjoint, interchanging the transitions leaves their four sites unchanged, so the factors in Eq.~(\ref{eqt:signed-M}) other than the signs are unchanged. For $\xi\in T_1$, performing $\tau_2$ first changes the parity of its index in decreasing order by $\#\{\zeta\in T_2:\zeta>\xi\}\ (\operatorname{mod}2)$, and similarly for the other transition. The sum of the changes in the two sign exponents is, modulo $2$,
\[
\sum_{\xi\in T_1}\#\{\zeta\in T_2:\zeta>\xi\}+\sum_{\zeta\in T_2}\#\{\xi\in T_1:\xi>\zeta\}=|T_1|\,|T_2|=16\equiv0\ (\operatorname{mod}2).
\]
Indeed, since $T_1\cap T_2=\varnothing$, each pair $(\xi,\,\zeta)\in T_1\times T_2$ is counted in exactly one of the two sums. Consequently,
\begin{equation}\label{eqt:path-numerator}
M_{\lambda\mu}M_{\mu\nu}=M_{\lambda\mu'}M_{\mu'\nu}.
\end{equation}

Next, compare the denominators. The disjoint supports imply that interchanging the transitions does not change their respective increments in $\mathsf p_3,\,\mathsf p_4$. Since $E_\lambda(P)=E_\nu(P)=E^{(0)}$, we obtain
\[
E_{\mu'}(P)-E^{(0)}=E_\nu(P)-E_\mu(P)=E^{(0)}-E_\mu(P),
\]
so the two denominators have opposite signs and equal magnitudes. Together with Eq.~(\ref{eqt:path-numerator}), this shows that the contributions of the original and reordered paths to $K_{\lambda\nu}(P)$ cancel:
\[
\frac{M_{\lambda\mu}M_{\mu\nu}} {E^{(0)}-E_\mu(P)} + \frac{M_{\lambda\mu'}M_{\mu'\nu}} {E^{(0)}-E_{\mu'}(P)} =0.
\]

This involution cancels all summands in pairs. Thus, for every $P\notin\Sigma_N$, $K_{\lambda\nu}(P)=0$. Since $K_{\lambda\nu}(P)$ is a rational function of $P$, it follows that
\[
K_{\lambda\nu}(P)\equiv0.
\]
\end{proof}
\subsection{Separation of the second-order corrections}\label{app:separation-section}
We prove that the second-order corrections $K_\lambda(P)$ associated with distinct partitions in the same equivalence class are distinct as rational functions of $P$. 

By Lemma~\ref{lem:second-diagonal}, the second-order effective operator $K$ is diagonal in the Schur basis. For each $\lambda\in\mathcal G$, the formula
\[
K_\lambda(P):=K_{\lambda\lambda}(P) = \sum_{\mu\notin\mathcal G} \frac{|M_{\mu\lambda}|^2}{E_\lambda(P)-E_\mu(P)}
\]
therefore gives the corresponding second-order correction. If $\mu$ is obtained by a nontrivial transition $\{\alpha,\,\beta\}\longrightarrow\{\gamma,\,\delta\}$, then Eqs.~(\ref{eqt:amplitude}) and (\ref{eqt:gap}) give
\[
|M_{\mu\lambda}|^2 = (\alpha-\beta)^2(\gamma-\delta)^2,\quad E_\lambda(P)-E_\mu(P) = \frac32 \bigl( (\alpha-\beta)^2-(\gamma-\delta)^2 \bigr) \left( P+\frac S2 \right).
\]
Grouping all transitions by $S$, we obtain
\begin{equation}\label{eqt:K}
K_\lambda(P)=\sum_{S\in\mathbb Z}\frac{R_\lambda(S)}{P+\frac{S}{2}},
\end{equation}
where
\begin{equation}\label{eqt:R}
R_\lambda(S)=\frac23\sum_{\substack{\alpha>\beta,\ \alpha,\,\beta\in\operatorname{Occ}(\lambda)\\
\gamma>\delta,\ \gamma,\,\delta\in\operatorname{Vac}(\lambda)\\
\alpha+\beta=\gamma+\delta=S
}}
\frac{(\alpha-\beta)^2(\gamma-\delta)^2}{(\alpha-\beta)^2-(\gamma-\delta)^2}.
\end{equation}
All Maya sites are half-integers, so $S\in\mathbb Z$, $R_\lambda(S)\in\mathbb Q$. Moreover, $\operatorname{Occ}(\lambda)$ has a maximum and $\operatorname{Vac}(\lambda)$ has a minimum. The conservation law $\alpha+\beta=\gamma+\delta$ therefore confines all admissible transitions to a finite interval. Hence, only finitely many $R_\lambda(S)$ are nonzero, and
\[
K_\lambda(P)\in\mathbb Q(P).
\]

To compare $K_\lambda$ for partitions in the same equivalence class, consider the Laurent expansion of $K_\lambda$ at $P=\infty$. Since $R_\lambda(S)$ is nonzero at only finitely many integers $S$, substituting $\frac1{P+\frac{S}{2}} = \sum_{m=0}^{\infty} \left(-\frac12\right)^m S^mP^{-m-1}$ into Eq.~(\ref{eqt:K}) gives
\begin{equation}\label{eqt:laurent}
K_\lambda(P)=\sum_{m=0}^{\infty}\left(-\frac12\right)^m\left(\sum_{S\in\mathbb Z}S^mR_\lambda(S)\right)P^{-m-1}.
\end{equation}
Denote the finite sums in these coefficients by
\begin{equation}\label{eqt:moment}
\mathcal M_m(\lambda):=\sum_{S\in\mathbb Z}S^mR_\lambda(S),
\end{equation}
where $m\ge0$. Thus, $K_\lambda(P)\equiv K_\mu(P)$ if and only if $\mathcal M_m(\lambda)=\mathcal M_m(\mu),\,\forall m\ge0$.

Assign the weights $\operatorname{wt}(\mathsf p_k)=k$, with the weight of any monomial defined as the sum of the weights of its factors. We will show that $\mathcal M_m$ is a polynomial in the shifted power sums, that each monomial has weight at most $m+5$, and that the coefficient of $\mathsf p_{m+5}$ is nonzero. Assuming these results, let $\lambda,\,\mu$ belong to the same equivalence class. If $K_\lambda(P)\equiv K_\mu(P)$, equivalently,
\[
\mathcal M_m(\lambda)=\mathcal M_m(\mu), \  \forall m\ge0,
\]
we can solve recursively to obtain $\mathsf p_{m+5}(\lambda)=\mathsf p_{m+5}(\mu),\,\forall m\ge0$. Newton's identities then give $\lambda=\mu$, meaning that distinct partitions in the same equivalence class have distinct second-order corrections $K_\lambda(P)$ as rational functions of $P$. 
\subsubsection{Rational polynomiality of the Laurent coefficients}
We establish the following Lemma \ref{lem:moment-theorem} in this subsubsection. 
\begin{lemma}\label{lem:moment-theorem}
For each integer $m\ge0$, the moment $\mathcal M_m$ is a polynomial in the shifted power sums with rational coefficients, satisfying
\[
\mathcal M_m\in\mathbb Q[\mathsf p_2,\,\cdots,\,\mathsf p_{m+5}], \quad \operatorname{wt}(\mathcal M_m)\le m+5, \quad \deg(\mathcal M_m)\le3.
\]
Here, $\deg$ denotes the ordinary total degree in the shifted power sums, and the polynomial coefficients are independent of the partition.
\end{lemma}

After translating the Maya diagram of any partition sufficiently far to the right by an integer, all negative half-integer sites are occupied, and only finitely many positive sites are occupied. We therefore first consider a finite set
\[
A\subset\left(\mathbb Z+\frac12\right)_{>0}, \quad \Omega_A := \left(\mathbb Z+\frac12\right)_{<0}\cup A,
\]
and define
\begin{equation}\label{eqt:F}
F_m(A):=\frac23\sum_{\substack{\alpha>\beta,\ \alpha,\,\beta\in\Omega_A\\\gamma>\delta,\ \gamma,\,\delta\notin\Omega_A\\\alpha+\beta=\gamma+\delta}}(\alpha+\beta)^m\frac{(\alpha-\beta)^2(\gamma-\delta)^2}{(\alpha-\beta)^2-(\gamma-\delta)^2}.
\end{equation}
Since $\Omega_A$ contains all negative half-integers and only finitely many positive occupied sites, the conservation law $\alpha+\beta=\gamma+\delta$ confines all admissible transitions to a finite interval. Thus, Eq.~(\ref{eqt:F}) is a finite sum. 

Fix a finite set $\mathcal U\subset(\mathbb Z+\frac12)_{>0}$, let $A\subset\mathcal U$, and let $n_\alpha\in\{0,\,1\}$ denote the occupation variable of $\alpha\in\mathcal U$. Then $F_m(A)$ has a unique expansion
\[
F_m(A) = \sum_{B\subset\mathcal U} C_{B,\,m}\prod_{\alpha\in B}n_\alpha,
\]
where
\begin{equation}\label{eqt:boolean-coeff}
C_{B,\,m}
=
\sum_{A\subseteq B}
(-1)^{|B|-|A|}F_m(A).
\end{equation}
For $|B|=1,\,2,\,3$, denote the corresponding coefficients by $C_{1,\,m}(\alpha),\,C_{2,\,m}(\alpha,\,\beta),\,C_{3,\,m}(\alpha,\,\beta,\,\gamma)$.

We first show that the expansion in occupation variables is at most cubic and bound the degrees of its coefficients. We then express it in ordinary power sums and use translation to obtain the weight bound in shifted power sums.
\begin{lemma}\label{lem:boolean-degree}
As a multilinear polynomial in the occupation variables, $F_m$ has degree at most $3$. Consequently,
\begin{equation}\label{eqt:boolean}
F_m(A)=\sum_{\alpha\in A}C_{1,\,m}(\alpha)+\sum_{\{\alpha,\,\beta\}\subset A}C_{2,\,m}(\alpha,\,\beta)+\sum_{\{\alpha,\,\beta,\,\gamma\}\subset A}C_{3,\,m}(\alpha,\,\beta,\,\gamma).
\end{equation}
\end{lemma}

\begin{proof}
If $|B|\ge5$, each transition involves at most four positive sites, so there is some $r\in B$ not involved in that transition. In Eq.~(\ref{eqt:boolean-coeff}), a term containing $r$ and the corresponding term not containing $r$ have equal contributions and opposite signs. Hence, $C_{B,\,m}=0$.

Suppose $|B|=4$. Only transitions involving all four sites in $B$ can contribute. For any such transition, the four sites split into $\{\alpha,\,\beta\}$ and $\{\gamma,\,\delta\}$, with $\alpha+\beta=\gamma+\delta$. Denote
\[
\kappa_m(\alpha,\,\beta;\,\gamma,\,\delta) := \frac23(\alpha+\beta)^m \frac{(\alpha-\beta)^2(\gamma-\delta)^2} {(\alpha-\beta)^2-(\gamma-\delta)^2}.
\]
The forward and reverse transitions contribute $\kappa_m(\alpha,\,\beta;\,\gamma,\,\delta)$ and $\kappa_m(\gamma,\,\delta;\,\alpha,\,\beta)$, respectively, and
\[
\kappa_m(\gamma,\,\delta;\,\alpha,\,\beta) = -\kappa_m(\alpha,\,\beta;\,\gamma,\,\delta).
\]
Thus, $C_{B,\,m}=0$ also holds for $|B|=4$. Finally, $F_m(\varnothing)=0$, so the constant term vanishes as well.
\end{proof}

\begin{lemma}\label{lem:boolean-coefficient-bounds}
The coefficients in Eq.~(\ref{eqt:boolean}) are polynomials with rational coefficients, and
\begin{equation}\label{eqt:coefficient-degrees}
\begin{aligned}
C_{1,\,m}&\in\mathbb Q[\alpha],
&
\deg C_{1,\,m}&\le m+4,\\
C_{2,\,m}&\in\mathbb Q[\alpha,\,\beta]^{\mathrm{S}_2},
&
\deg C_{2,\,m}&\le m+3,\\
C_{3,\,m}&\in\mathbb Q[\alpha,\,\beta,\,\gamma]^{\mathrm{S}_3},
&
\deg C_{3,\,m}&\le m+2.
\end{aligned}
\end{equation}
\end{lemma}

\begin{proof}
We consider the three types of coefficients separately.

\noindent \textbf{(i) Three-site coefficients.}

Fix distinct positive sites $\alpha,\,\beta,\,\gamma$. By Eq.~(\ref{eqt:boolean-coeff}), $C_{3,\,m}(\alpha,\,\beta,\,\gamma)$ is the coefficient of $n_\alpha n_\beta n_\gamma$ in the occupation polynomial. Setting the occupation variables at all other positive sites to zero leaves this coefficient unchanged.

First consider the pairs $\{\alpha,\,\beta\}$ and $\{\gamma,\,\delta\}$, where $\delta=\alpha+\beta-\gamma$. Distinctness gives $\delta\ne\alpha,\,\beta$; assume for now that $\delta\ne\gamma$. If $\delta>0$, the reverse transition has weight $-\kappa_m(\alpha,\,\beta;\,\gamma,\,\delta)$, so the total contribution of the two directions is
\[
\kappa_m(\alpha,\,\beta;\,\gamma,\,\delta)\bigl[n_\alpha n_\beta(1-n_\gamma)(1-n_\delta)-n_\gamma n_\delta(1-n_\alpha)(1-n_\beta)\bigr].
\]
To extract the required coefficient, set $n_\delta=0$. If $\delta<0$, this site is always occupied, and only the transition from $\{\gamma,\,\delta\}$ to $\{\alpha,\,\beta\}$ can contribute. Its occupation polynomial is
\[
-\kappa_m(\alpha,\,\beta;\,\gamma,\,\delta)n_\gamma(1-n_\alpha)(1-n_\beta).
\]
In both cases, the coefficient of $n_\alpha n_\beta n_\gamma$ is $-\kappa_m(\alpha,\,\beta;\,\gamma,\,\delta)$. Since $(\alpha-\beta)^2-(\alpha+\beta-2\gamma)^2=-4(\gamma-\alpha)(\gamma-\beta)$, substituting into $-\kappa_m$ and summing over cyclic permutations of $(\alpha,\,\beta,\,\gamma)$ gives
\begin{equation}\label{eqt:C3}
C_{3,\,m}(\alpha,\,\beta,\,\gamma)=\frac16\sum_{\mathrm{cyc}}\frac{(\alpha-\beta)^2(\alpha+\beta-2\gamma)^2(\alpha+\beta)^m}{(\gamma-\alpha)(\gamma-\beta)}.
\end{equation}
If $\delta=\gamma$, then $\alpha+\beta-2\gamma=0$, so the corresponding term vanishes and the formula remains valid.

We next show that the denominators in Eq.~(\ref{eqt:C3}) cancel. Denote
\[
\Gamma(\alpha,\,\beta,\,\gamma):=(\alpha-\beta)(\alpha-\gamma)(\beta-\gamma).
\]
By Eq.~(\ref{eqt:C3}), $\Gamma C_{3,\,m}$ is a polynomial with rational coefficients of degree at most $m+5$. By definition, $C_{3,\,m}$ is symmetric in its three variables, so $\Gamma C_{3,\,m}$ is alternating and hence divisible by $\Gamma$. Dividing by the degree-$3$ factor $\Gamma$ gives
\begin{equation}\label{eqt:degree3}
C_{3,\,m}\in\mathbb Q[\alpha,\,\beta,\,\gamma]^{\mathrm{S}_3}, \quad \deg C_{3,\,m}\le m+2.
\end{equation}

\noindent \textbf{(ii) Two-site coefficients.}

First set $m=0$, so the factor involving the sum of the sites in Eq.~(\ref{eqt:F}) is $1$. For an integer $L\ge0$, denote
\[
I_L:=\left\{\frac12,\,\frac32,\,\cdots,\,L-\frac12\right\}.
\]
For $A\subset I_L$, apply reflection followed by complementation:
\[
A^\vee:=I_L\setminus(L-A), \ n_\gamma(A^\vee)=1-n_{L-\gamma}(A), \  \gamma\in I_L.
\]
This maps the transition $\{\alpha,\,\beta\}\longrightarrow\{\gamma,\,\delta\}$ to $\{L-\gamma,L-\delta\}\longrightarrow\{L-\alpha,L-\beta\}$. The squared differences on the two sides are interchanged, leaving the numerator unchanged and reversing the sign of the denominator. Hence,
\begin{equation}\label{eqt:F0-duality}
F_0(A^\vee)=-F_0(A).
\end{equation}
Fix $\alpha>\beta>0$ and take $L=\alpha+\beta$, so reflection interchanges $\alpha,\,\beta$. Make the above substitution of occupation variables in Eq.~(\ref{eqt:boolean}). To extract $n_\alpha n_\beta$, take the negative term from each of the two relevant factors in a quadratic term; in a cubic term, also take the constant term from the remaining factor. The coefficient in $F_0(A^\vee)$ is therefore $C_{2,\,0}(\alpha,\,\beta)+\sum_{\gamma\in I_{\alpha+\beta}\setminus\{\alpha,\,\beta\}}C_{3,\,0}(\alpha,\,\beta,\,\gamma)$. The corresponding coefficient on the right-hand side of Eq.~(\ref{eqt:F0-duality}) is $-C_{2,\,0}(\alpha,\,\beta)$. Comparing the two gives
\begin{equation}\label{eqt:C20}
2C_{2,\,0}(\alpha,\,\beta)=-\sum_{\gamma\in I_{\alpha+\beta}\setminus\{\alpha,\,\beta\}}C_{3,\,0}(\alpha,\,\beta,\,\gamma).
\end{equation}
By Eq.~(\ref{eqt:degree3}) and Faulhaber's formula, the right-hand side is a symmetric polynomial of degree at most $3$, and so is $C_{2,\,0}$.

For general $m$, multiplying every contribution for $m=0$ by $S^m$ gives $S^mC_{2,\,0}(\alpha,\,\beta)$. If $\alpha,\,\beta$ are both source sites or both target sites, their conserved sum is exactly $S$. If one is a source and the other a target, the weight must be corrected using the actual conserved sum.

When extracting the coefficient of $n_\alpha n_\beta$, all other positive sites are vacant and all negative sites are occupied. For a transition of the latter type, the other source site is negative and the other target site is positive. Conservation of the sum therefore forces the positive source to exceed the positive target, so $\alpha$ is the source and $\beta$ the target. The transition has the unique form
\[
\{\alpha,\,\beta-t\}\longrightarrow\{\beta,\,\alpha-t\}, \  t\in\mathbb Z, \  \beta+\frac12\le t\le\alpha-\frac12.
\]
The conditions $\beta-t<0$ and $\alpha-t>0$ require the other source to be a negative occupied site and the other target to be a positive vacant site, respectively. At $t=x$, the targets coincide and the contribution vanishes. The conserved sum is $S-t$, and $\bigl(\alpha-(\beta-t)\bigr)^2-\bigl((\alpha-t)-\beta\bigr)^2=(x+t)^2-(x-t)^2=4xt$, so the weight in Eq.~(\ref{eqt:F}) is $\frac{(t^2-x^2)^2}{6xt}(S-t)^m$. The occupation factor $n_\alpha(1-n_\beta)$ contributes a minus sign. Subtracting the reference weight $S^m$ from the actual weight $(S-t)^m$ and summing over $t$ gives
\begin{equation}\label{eqt:C2}
C_{2,\,m}(\alpha,\,\beta)=S^mC_{2,\,0}(\alpha,\,\beta)-\frac1{6x}\sum_{t=\beta+\frac{1}{2}}^{\alpha-\frac{1}{2}}(t^2-x^2)^2\frac{(S-t)^m-S^m}{t}.
\end{equation}
For $m\ge1$, the quotient $\frac{(S-t)^m-S^m}{t}$ is a polynomial of degree $m-1$, so the summand has total degree at most $m+3$. By Faulhaber's formula, the finite sum in Eq.~(\ref{eqt:C2}), before division by $x$, is a polynomial in $(\alpha,\,\beta)$ of degree at most $m+4$. It is the difference of polynomial values at the endpoints $\alpha-\frac12$ and $\beta-\frac12$, so it vanishes at $\alpha=\beta$. It is therefore divisible by $x=\alpha-\beta$, with quotient of degree at most $m+3$.

Interchanging $\alpha,\,\beta$ leaves $S,\,x^2$ unchanged and exchanges the endpoints. Thus, the finite sum and $x$ both change sign, so their quotient is symmetric. Together with $\deg(S^mC_{2,\,0})\le m+3$, this gives
\begin{equation}\label{eqt:degree2}
C_{2,\,m}\in\mathbb Q[\alpha,\,\beta]^{\mathrm{S}_2}, \quad\deg C_{2,\,m}\le m+3.
\end{equation}

\noindent \textbf{(iii) One-site coefficients.}

Let
\[
I_n = \left\{ \frac12,\,\frac32,\,\cdots,\,n-\frac12 \right\}.
\]
Every occupied site in $\Omega_{I_n}$ is smaller than every vacant site, so no transition can conserve the sum of the sites. Hence, $F_m(I_n)=0$. Taking the difference $F_m(I_{n+1})-F_m(I_n)=0$ and using Eq.~(\ref{eqt:boolean}), we obtain
\begin{equation}\label{eqt:C1}
C_{1,\,m}\left(n+\frac12\right)=-\sum_{j=0}^{n-1}C_{2,\,m}\left(n+\frac12,\,j+\frac12\right)-\sum_{0\le j<k<n}C_{3,\,m}\left(n+\frac12,\,j+\frac12,\,k+\frac12\right).
\end{equation}
By Eqs.~(\ref{eqt:degree3}) and (\ref{eqt:degree2}), and Faulhaber's formula, the right-hand side is a polynomial in $n$ of degree at most $m+4$. Thus, 
\begin{equation}\label{eqt:degree1}
C_{1,\,m}\in\mathbb Q[\alpha],\quad\deg C_{1,\,m}\le m+4.
\end{equation}
\end{proof}
We now express the sums of these coefficient polynomials over distinct sites in terms of ordinary power sums. 
\begin{lemma}\label{lem:finite-polynomiality}
Denote
\[
\pi_j(A) := \sum_{\alpha\in A}\alpha^j, \quad \pi_0(A):=|A|, \quad \operatorname{wt}(\pi_j):=j+1.
\]
There exists a polynomial $\Psi_m$ with rational coefficients, depending only on $m$, such that
\begin{equation}\label{eqt:Psi}
F_m(A)
=
\Psi_m
\bigl(
\pi_0(A),\,\pi_1(A),\,\cdots
\bigr),
\quad
\operatorname{wt}(\Psi_m)\le m+5,
\end{equation}
and each monomial of $\Psi_m$ contains at most three ordinary power-sum factors.
\end{lemma}

\begin{proof}
By Lemmas~\ref{lem:boolean-degree} and \ref{lem:boolean-coefficient-bounds}, Eq.~(\ref{eqt:boolean}) contains only sums over $r=1,\,2,\,3$ sites. Since the coefficients are symmetric, each unordered sum equals $\frac{1}{r!}$ times the corresponding ordered sum. Expand the coefficient polynomials into monomials and abbreviate $\pi_j=\pi_j(A)$. For $u,\,v,\,w\ge0$, we have
\[
\sum_{\alpha\in A}\alpha^u=\pi_u, \quad \sum_{\substack{\alpha,\,\beta\in A\\\alpha\ne\beta}}\alpha^u\beta^v=\pi_u\pi_v-\pi_{u+v},
\]
and
\[
\sum_{\substack{\alpha,\,\beta,\,\gamma\in A\\\text{pairwise distinct}}}\alpha^u\beta^v\gamma^w=\pi_u\pi_v\pi_w-\pi_{u+v}\pi_w-\pi_{u+w}\pi_v-\pi_{v+w}\pi_u+2\pi_{u+v+w}.
\]
The last term restores twice the contribution from triples with all three sites equal.

These conversions only merge power-sum indices. For a monomial of total degree $d$ in $r$ sites, every resulting term $\pi_{j_1}\cdots\pi_{j_s}$ satisfies $s\le r$ and $j_1+\cdots+j_s=d$. By Eqs.~(\ref{eqt:coefficient-degrees}), $d\le m+5-r$, so
\[
\operatorname{wt}(\pi_{j_1}\cdots\pi_{j_s})=d+s\le d+r\le m+5.
\]
Thus, each term contains at most three power-sum factors and has weight at most $m+5$. The resulting polynomial has rational coefficients independent of $A$ and depending only on $m$, as required.
\end{proof}

We now transfer the result for finite configurations to arbitrary partitions.
\begin{lemma}\label{lem:translation-lemma}
Let $\lambda$ be any partition, let $L\ge\ell(\lambda)$, where $\ell(\lambda)$ is used to denote the length of $\lambda$, and denote
\[
A_L := \left\{ \lambda_i+L-i+\frac12: 1\le i\le L \right\}.
\]
Then
\begin{equation}\label{eqt:translate}
\mathcal M_m(\lambda)
=
\sum_{k=0}^m
\binom mk
(-2L)^{m-k}F_k(A_L),
\end{equation}
and, for $j\ge1$,
\begin{equation}\label{eqt:rawshift}
\pi_j(A_L)
=
\sum_{\ell=1}^j
\binom j\ell
L^{j-\ell}\mathsf p_{\ell+1}(\lambda)
+
\sum_{i=1}^L
\left(
L-i+\frac12
\right)^j,
\end{equation}
while $\pi_0(A_L)=L$.
\end{lemma}

\begin{proof}
Translating the Maya diagram of $\lambda$ to the right by $L$ gives the occupied set $\left(\mathbb Z+\frac12\right)_{<0}\cup A_L$. Translation preserves all site differences in a transition and increases the conserved sum by $2L$. Binomial inversion applied to $(S+2L)^k$ gives Eq.~(\ref{eqt:translate}). Eq.~(\ref{eqt:rawshift}) follows directly from the definition of $A_L$ and the shifted power sums.
\end{proof}

\begin{proof}[Proof of Lemma~\ref{lem:moment-theorem}]
By Lemmas~\ref{lem:finite-polynomiality} and \ref{lem:translation-lemma}, the right-hand side of Eq.~(\ref{eqt:translate}) is a polynomial in $L$. For every integer $L\ge\ell(\lambda)$, this polynomial equals $\mathcal M_m(\lambda)$, so it is constant. We may therefore evaluate it at $L=0$. Only the term with
$k=m$ survives in Eq.~(\ref{eqt:translate}), while Eq.~(\ref{eqt:rawshift}) gives the formal substitution
\[
\pi_0\longmapsto0, \ \pi_j\longmapsto\mathsf p_{j+1}(\lambda), \ j\ge1.
\]
Hence,
\begin{equation}\label{eqt:momentproved}
\mathcal M_m(\lambda)
=
\Psi_m
\bigl(
0,\,
\mathsf p_2(\lambda),\,
\mathsf p_3(\lambda),\,
\cdots
\bigr).
\end{equation}
Since $\pi_j$ and $\mathsf p_{j+1}$ have the same weight $j+1$, both the weight bound and the bound on the number of power-sum factors in Eq.~(\ref{eqt:Psi}) are preserved. This proves the lemma.
\end{proof}
\subsubsection{Spectral separation of the second-order effective operator}
By the weight bound in Lemma~\ref{lem:moment-theorem}, $\mathsf p_{m+5}$ can occur only as a linear term with no other factors. Thus,
\begin{equation}\label{eqt:triangular}
\mathcal M_{m}=\alpha_m\mathsf p_{m+5}
+\Theta_m(\mathsf p_{2},\,\cdots,\,\mathsf p_{m+4}),  \ \alpha_m\in\mathbb Q.
\end{equation}
Here, $\Theta_m$ is a polynomial in the lower shifted power sums. To recover the shifted power sums recursively, it suffices to prove that $\alpha_m$ is nonzero. One-row partitions provide a positive lower bound for this coefficient.

\begin{lemma}\label{eqt:alpha-positive}
For every integer $m\ge0$, we have $\alpha_m>0$.
\end{lemma}
\begin{proof}
Consider the one-row partition $(n)$. Then $\mathsf p_k((n))=\left(n-\frac12\right)^{k-1}-\left(-\frac12\right)^{k-1}$, so $\mathsf p_k((n))$ is a monic polynomial of degree $k-1$. Consequently, 
\[
\mathsf p_{m+5}((n))=n^{m+4}+O(n^{m+3}).
\]
Every other monomial in Eq.~(\ref{eqt:triangular}) either consists of a single $\mathsf p_k$ with $k\le m+4$ or contains at least two factors. The weight bound $m+5$ implies that each has degree at most $m+3$ in $n$. Hence,
\begin{equation}\label{eqt:alpha-limit}
\mathcal M_m((n))=\alpha_m n^{m+4}+O(n^{m+3}).
\end{equation}

A one-row partition has exactly one positive occupied site. By Eq.~(\ref{eqt:R}), the source and target pairs can be parametrized as
\[
(\alpha,\,\beta)=\left(n-\frac12,\,-j-\frac12\right), \quad (\gamma,\,\delta)=\left(n-j-k-\frac12,\,k-\frac12\right).
\]
Therefore,
\begin{equation}\label{eqt:row}
\mathcal M_{m}((n))=\sum_{j=1}^{n-1}\sum_{0\le k<\frac{(n-j)}{2}}
\frac{(n-j-1)^m(n+j)^2(n-j-2k)^2}{6(j+k)(n-k)}.
\end{equation}
All summands are nonnegative. Let $n$ be a positive multiple of $4$ and retain only the terms with $1\le j,\,k\le \frac{n}{4}$. This range satisfies $2k<n-j$, and
\[
n-j-1\ge \frac{n}{2},  \quad n+j\ge n,  \quad n-j-2k\ge \frac{n}{4},  \quad 6(j+k)(n-k)\le3n^2.
\]
It follows that
\[
\mathcal M_{m}((n))\ge\sum_{j,\,k=1}^{\frac{n}{4}}\frac{(\frac{n}{2})^m n^2(\frac{n}{4})^2}{3n^2}=\frac{n^{m+4}}{768\cdot2^m}.
\]
Divide by $n^{m+4}$ and let $n\to+\infty$. By Eq.~(\ref{eqt:alpha-limit}), we obtain
\[
\alpha_m\ge(768\cdot2^m)^{-1}>0.
\]
\end{proof}

\begin{lemma}\label{lem:separation}
Let $\lambda,\,\mu$ be two partitions at the same level. If 
\[
\mathsf p_3(\lambda)=\mathsf p_3(\mu), \quad \mathsf p_4(\lambda)=\mathsf p_4(\mu), \quad K_\lambda(P)\equiv K_\mu(P),
\]
then $\lambda=\mu$.
\end{lemma}

\begin{proof}
By Eq.~(\ref{eqt:laurent}) and $K_\lambda(P)\equiv K_\mu(P)$, comparison of Laurent coefficients at $P=\infty$ gives $\mathcal M_m(\lambda)=\mathcal M_m(\mu),\,m\ge0$. Since $\lambda,\,\mu$ belong to the same level and equivalence class, we also have $\mathsf p_2(\lambda)=\mathsf p_2(\mu),\ \mathsf p_3(\lambda)=\mathsf p_3(\mu), \ \mathsf p_4(\lambda)=\mathsf p_4(\mu)$. Using the triangular relation
\[
\mathcal M_m = \alpha_m\mathsf p_{m+5} + \Theta_m(\mathsf p_2,\,\cdots,\,\mathsf p_{m+4}), \ \alpha_m\neq0,
\]
and induction on $m$, we obtain
\begin{equation}\label{eqt:all-shifted-powers}
\mathsf p_k(\lambda)=\mathsf p_k(\mu),\ k\ge2.
\end{equation}

Choose
\[
L\ge\max\{\ell(\lambda),\,\ell(\mu)\}, \ a_i:=\lambda_i-i+\frac12, \ b_i:=\mu_i-i+\frac12, \ 1\le i\le L.
\]
By Eq.~(\ref{eqt:all-shifted-powers}), cancelling the common vacuum terms gives
\[
\sum_{i=1}^L a_i^j = \sum_{i=1}^L b_i^j, \ 1\le j\le L.
\]
Newton's identities~\cite[Proposition~2.6.3]{Savage:2022gtg} yield
\[
\prod_{i=1}^L(t-a_i) = \prod_{i=1}^L(t-b_i).
\]
Thus, the two multisets of roots coincide. Since both $a_i$ and $b_i$ are strictly decreasing, we have $a_i=b_i$ for all $1\le i\le L$, and hence
$\lambda=\mu$.
\end{proof}

At fixed level $N\ge4$, it suffices that $\mathcal M_0,\,\cdots,\,\mathcal M_{N-4}$ agree for $\lambda,\,\nu$: one then successively determines $\mathsf p_2,\,\mathsf p_3,\,\mathsf p_4,\,\mathsf p_5,\,\cdots,\,\mathsf p_{N+1}$, and Newton's identities show that the partitions coincide. For $N\le3$, equality of $\mathsf p_2,\,\mathsf p_3,\,\mathsf p_4$ already determines the partition.

\begin{proof}[Proof of Lemma~\ref{lem:3}]
Fix $N\ge0$ and any real transcendental number $P_*$. Since $\Sigma_N\subset\mathbb Q$, we have $P_*\notin\Sigma_N$. Thus, the unperturbed eigenvalues $E_{\mathcal G}(P_*)$ associated with distinct full eigenspaces are pairwise distinct.

For any equivalence class $\mathcal G$ and distinct $\lambda,\mu\in\mathcal G$, Lemma~\ref{lem:separation} gives $K_\lambda(P)-K_\mu(P)\not\equiv0$. Since $K_\lambda-K_\mu\in\mathbb Q(P)$ and $P_*$ is transcendental, it follows that
\[
K_\lambda(P_*)\neq K_\mu(P_*).
\]
Second-order perturbation therefore separates the eigenvalues within each full eigenspace for all sufficiently small nonzero $s$, while the gaps between distinct full eigenspaces persist by continuity. Since there are finitely many partitions, there exists $\varepsilon_N(P_*)>0$ such that $\widehat D_N(P_*,\,s)$ has simple spectrum whenever $0<|s|<\varepsilon_N(P_*)$, and so does the corresponding $D_N$. Returning to the original parameters gives
\[
\operatorname{disc}_N \left( -2+6s,\, \frac{P_*^2}{2}-\frac18+\frac s4 \right) \neq0.
\]
Hence,
\[
\operatorname{disc}_N(c,\,\Delta)\not\equiv0.
\]
\end{proof}

In conclusion, the measure of the degenerate region of the first two qKdV charges is zero, and Theorem \ref{theo:1} is proved.
\section{Solution of microstates with qKdV charges}\label{app:sol}
This appendix focuses on the solution of microstates with qKdV charges. We first derive the formula of $\{n_k\}$ in terms of $\{M_{2r-1}\}$ in Appendix \ref{app:reconst}, and then obtain the formula of $\{n_k\}$ and $M_{0,\,2}$ in $\{q_{2r-1}\}$ in Appendix \ref{app:inversion} by substituting $\{M_{2r-1}\}$ with $\{q_{2r-1}\}$. At the end of Appendix \ref{app:inversion}, we clarify the technique of analytic continuation to highlight physical meaning of the formulae. In Appendix \ref{app:constraint}, we derive additional constraints on the highest weight imposed by the inversion formula Eq.~(\ref{eqt:reconst}).

To ensure the mathematical rigor of the analytic continuation, we do not take the thermodynamic limit $\ell\to+\infty$ in any of the derivations in this appendix unless we directly discuss the behaviors in that limit. Keep in mind in this appendix that for physical systems under our discussion, the finiteness of the Hamiltonian $Q_1$ renders 
\[
\exists k_0\in\mathbb{N},\ \forall k\geq k_0,\ n_k=0.
\]
\subsection{Proof of the reconstruction formula}\label{app:reconst}
Here, we prove the reconstruction formula Eq.~(\ref{eqt:reconst}), i.e.~express $\{n_k\}$ in terms of $\{M_{2r-1}\}$.

We start from
\begin{equation}
M_{2p+1}=\sum_{j=1}^{\infty} (jn_j)(j^2)^p .
\end{equation}
We define
\begin{equation}
S(x)=\frac{\sin(\pi\sqrt{x})}{\pi\sqrt{x}}=\sum_{m=0}^{\infty}\frac{(-1)^m\pi^{2m}}{(2m+1)!}x^m,
\end{equation}
and then
\begin{equation}
S(0)=1,\ S(k^2)=0,\ \forall k\in\mathbb{N}_+.
\end{equation}
Note that
\begin{equation}
\lim_{x\to k^2} \frac{S(x)}{x-k^2}=S'(k^2)=\frac{(-1)^k}{2k^2}.
\end{equation}
Then one defines Eq.~(\ref{eqt:app11}):
\begin{align}
\Phi_k(x)&=2(-1)^k k^2\frac{\sin(\pi\sqrt{x})}{\pi\sqrt{x}(x-k^2)}=-2(-1)^k\frac{\sin(\pi\sqrt{x})}{\pi\sqrt{x}}\frac{1}{1-\frac{x}{k^2}}\label{eqt:app11}\\&=
\begin{cases}
\displaystyle-2(-1)^k \left[\sum_{m=0}^{\infty}\frac{(-1)^m\pi^{2m}}{(2m+1)!}x^m\right]\left[\sum_{q=0}^{\infty}\frac{x^q}{k^{2q}}\right],\ &x<k^2,\\[0.1em]
\displaystyle-2(-1)^k \left[\sum_{m=0}^{\infty}\frac{(-1)^m\pi^{2m}}{(2m+1)!}x^m\right]\left[\sum_{q=1}^{\infty}\frac{k^{2q}}{x^{q}}\right],\ &x>k^2,
\end{cases}\label{eqt:app12}
\end{align}
so that for $j\in\mathbb{N}_+$,
\begin{equation}
\Phi_k(j^2)=\delta_{kj}.
\end{equation}
We can expand Eq.~(\ref{eqt:app11}) with $x$:
\begin{equation}\label{eqt:app21}
\Phi_k(x)=\sum_{p=0}^{\infty}A_{kp}x^p,
\end{equation}
where for $x<k^2$, 
\begin{equation}\label{eqt:app31}
A_{kp}=\frac{\Phi_k^{(p)}(0)}{p!}=-2(-1)^k\sum_{\substack{m,\,q=0\\m+q=p}}^{\infty}\frac{(-1)^m\pi^{2m}}{(2m+1)!}\frac{1}{k^{2q}}=-2(-1)^k k^{-2p}\sum_{m=0}^{p}\frac{(-1)^m(\pi k)^{2m}}{(2m+1)!}.
\end{equation}
The result is the same for $x>k^2$:
\begin{equation}\label{eqt:app32}
A_{kp}=\frac{\Phi_k^{(p)}(0)}{p!}=2(-1)^k k^{-2p}\sum_{m=p+1}^{\infty}\frac{(-1)^m(\pi k)^{2m}}{(2m+1)!}=-2(-1)^k k^{-2p}\sum_{m=0}^{p}\frac{(-1)^m(\pi k)^{2m}}{(2m+1)!},
\end{equation}
due to absolute convergence. Meanwhile,
\begin{equation}\label{eqt:app3}
\sum_{j=1}^{\infty} jn_j\Phi_k(j^2)=kn_k .
\end{equation}
Combining Eqs.~(\ref{eqt:app21}), (\ref{eqt:app31}), (\ref{eqt:app32}) and (\ref{eqt:app3}), and letting $r=p+1$, we obtain Eq.~(\ref{eqt:reconst}):
\[
n_k=-2(-1)^k \sum_{r=1}^{\infty}\sum_{m=0}^{r-1}\frac{(-1)^m(\pi k)^{2m}}{(2m+1)!}\frac{M_{2r-1}}{k^{2r-1}} .
\]
\subsection{Inversion from qKdV charges to occupation numbers}\label{app:inversion}
This subappendix gives the explicit inversion formula from qKdV charges to occupation numbers and the expression of $M_{0,\,2}$. 
\subsubsection{Finite triangular inversion at fixed $h$}
Ref.~\cite{Dymarsky:2022dhi} gives in the $1/c$ expansion at fixed $h$ that
\begin{equation}
Q_{2r-1}=\tilde{\Delta}^r+\sum_{j=1}^r\frac{(2r-1)\sqrt\pi\,\Gamma(r+1)}{2\,\Gamma(j+\frac12)\Gamma(r-j+1)}\tilde{\Delta}^{r-j}\tilde{c}^{j-1}\left(M_{2j-1}+\frac12\zeta(1-2j)\right)+O(\tilde{c}^{r-2}).
\end{equation}
Accordingly, the exact finite-$\ell$ result recovered from Eq.~(\ref{eqt:m2q}) becomes, $\forall r\geq 2$, 
\begin{equation}\label{eqt:sum}
\tilde c\,\delta q_{2r-1}=2r(r-1)\frac{M_1-\frac{1}{24}}{h}+\sum_{j=2}^r\frac{(2r-1)\sqrt\pi\,\Gamma(r+1)}{2\,\Gamma(j+\frac12)\Gamma(r-j+1)}\frac{M_{2j-1}+\frac12\zeta(1-2j)}{h^j}+O(\tilde{c}^{-1}).
\end{equation}
Applying
\[
\Gamma\left(j+\frac12\right)=\frac{(2j)!}{4^j j!}\sqrt\pi,
\]
and including the $j=1$ term in Eq.~(\ref{eqt:sum}), we obtain
\begin{equation}\label{eqt:binomial-forward}
\frac{\tilde c\,\delta q_{2r-1}+r\frac{M_1-\frac1{24}}h}{2r-1}=\sum_{j=1}^r\binom rj\frac{4^j}{2\binom{2j}j}\frac{M_{2j-1}+\frac12\zeta(1-2j)}{h^j}+O(\tilde{c}^{-1}).
\end{equation}
\begin{lemma}\label{lem:inv}
If two sequences satisfy $b_r=\sum_{j=0}^r\binom rj a_j$, then
\[
a_r=\sum_{j=0}^r(-1)^{r-j}\binom rj b_j.
\]
\end{lemma}
\begin{proof}
Substituting the finite sum defining $b_j$ into the right-hand side and using
$\binom rj\binom js=\binom rs\binom{r-s}{j-s}$, we obtain
\[
\sum_{j=0}^r(-1)^{r-j}\binom rj b_j
=\sum_{s=0}^r\binom rs a_s\sum_{j=s}^r(-1)^{r-j}\binom{r-s}{j-s}.
\]
The innermost sum equals $(1-1)^{r-s}$, which vanishes for $s<r$ and equals one for $s=r$. Thus only $a_r$ remains. All sums involved are finite.
\end{proof}
Applying Lemma \ref{lem:inv} to Eq.~(\ref{eqt:binomial-forward}), we obtain
\begin{align}
&\phantom{=}\frac{4^r}{2\binom{2r}r}\frac{M_{2r-1}+\frac12\zeta(1-2r)}{h^r}\nonumber\\
&=\tilde{c}\,\sum_{j=2}^r(-1)^{r-j}\binom rj\frac{\delta q_{2j-1}}{2j-1}+\frac{M_1-\frac{1}{24}}{h}\sum_{j=1}^r(-1)^{r-j}\binom rj\frac{j}{2j-1}+O(\tilde{c}^{-1})\nonumber\\
&=\tilde{c}\,\sum_{j=2}^r(-1)^{r-j}\binom rj\frac{\delta q_{2j-1}}{2j-1}+\frac{M_1-\frac{1}{24}}{h}(-1)^{r-1}\frac{4^r}{2\binom{2r}r}+O(\tilde{c}^{-1}).
\end{align}
Therefore, $\forall r \geq 2$,
\begin{equation}\label{eqt:moment-inverse}
M_{2r-1}=(-h)^{r-1}\left(M_1-\frac1{24}\right)-\frac12\zeta(1-2r)+\tilde c\,h^r\frac{2\binom{2r}r}{4^r}\sum_{j=2}^r(-1)^{r-j}\binom rj\frac{\delta q_{2j-1}}{2j-1}+O(\tilde{c}^{-1}).
\end{equation}

Substituting Eq.~(\ref{eqt:moment-inverse}) into Eq.~(\ref{eqt:reconst}) gives Eq.~(\ref{eqt:inversion}): 
\[
\begin{aligned}
n_k&=-2(-1)^k\sum_{r=1}^{\infty}\frac{1}{k^{2r-1}}\left[\sum_{m=0}^{r-1}\frac{(-1)^m(\pi k)^{2m}}{(2m+1)!}\right]\left[\left(-\frac{\tilde{\Delta}}{\tilde{c}}\right)^{r-1}\left(\ell^2q_1-\tilde\Delta\right)\right.\\
&\phantom{=}\ \left.-\frac12\zeta(1-2r)+2\tilde{c}\,\binom{2r}{r}\left(\frac{\tilde{\Delta}}{4\tilde{c}}\right)^r\sum_{j=2}^{r}(-1)^{r-j}\binom{r}{j}\frac{\delta q_{2j-1}}{2j-1}\right]+O(\tilde{c}^{-1}).
\end{aligned}
\]
\subsubsection{Final result with analytic continuation}
In this part, we first work at large but finite $\ell$, and finally discuss the convergence issue of these formulae and the analytic continuation in the thermodynamic limit.

To recover the occupation numbers from the odd moments, define
\begin{equation}\label{eqt:F2}
F(x)=\sum_{j=1}^{\infty}n_j\sin(jx).
\end{equation}
Since the sum is finite, $F$ is entire. Expanding each sine function gives
\begin{equation}\label{eqt:F-Taylor}
F(x)=\sum_{r=1}^{\infty}\frac{(-1)^{r-1}M_{2r-1}}{(2r-1)!}x^{2r-1}.
\end{equation}
This Taylor series converges absolutely and uniformly on $[0,\pi]$ for any finite $\ell$, so it may be integrated term by term. The orthogonality relation 
\[
\int_0^\pi\sin(jx)\sin(kx)\,\mathrm{d}x=\frac\pi2\delta_{jk}
\]
then yields
\begin{equation}\label{eqt:on}
\begin{aligned}
n_k
&=\frac{2}{\pi}\int_0^\pi F(x)\sin(kx)\,\mathrm{d}x
=\frac{2}{\pi}\int_0^\pi
\left(\sum_{r=1}^{\infty}\frac{(-1)^{r-1}M_{2r-1}}{(2r-1)!}\,x^{2r-1}\right)\sin(kx)\,\mathrm{d}x\\
&=\frac{2}{\pi}\sum_{r=1}^{\infty}
\frac{(-1)^{r-1}M_{2r-1}}{(2r-1)!}
\int_0^\pi x^{2r-1}\sin(kx)\,\mathrm{d}x\\
&=\frac{2}{\pi}\sum_{r=1}^{\infty}
\frac{(-1)^{r-1}}{(2r-1)!}\left[\left(-\frac{\tilde{\Delta}}{\tilde{c}}\right)^{r-1}\left(\ell^2q_1-\tilde{\Delta}\right)-\frac12\zeta(1-2r)\phantom{\sum_{j=2}^r}\right.\\
&\left.\phantom{=}+\frac{\tilde{\Delta}^r}{\tilde c^{r-1}}\frac{2\binom{2r}r}{4^r}\sum_{j=2}^r(-1)^{r-j}\binom rj\frac{\delta q_{2j-1}}{2j-1}\right]
\int_0^\pi x^{2r-1}\sin(kx)\,\mathrm{d}x+O(\tilde{c}^{-1}).
\end{aligned}
\end{equation}
Repeated integration by parts gives Eq.~(\ref{eqt:reconst}) recovered.

The same generating function also gives an integral representation of the remaining term $M_{0,\,2}$ in the expressions of quasi-primary expectations at levels 6 and 8. Squaring Eq.~(\ref{eqt:F2}) and using orthogonality, we obtain
\begin{equation}\label{eqt:mf}
M_{0,\,2}=\sum_{k=1}^{\infty}n_k^2=\frac2\pi\int_0^\pi F(x)^2\,\mathrm{d}x.
\end{equation}

With the empty-sum convention, Eq.~(\ref{eqt:moment-inverse}) also holds for $r=1$. Substituting it into Eq.~(\ref{eqt:F-Taylor}) and working coefficient by coefficient in the Taylor series gives
\[
\begin{aligned}
F(x)
&=\sum_{r=1}^{\infty}\frac{(-1)^{r-1}M_{2r-1}}{(2r-1)!}x^{2r-1}\\
&=\sum_{r=1}^{\infty}\frac{(-1)^{r-1}x^{2r-1}}{(2r-1)!}
\left[(-h)^{r-1}\left(M_1-\frac1{24}\right)-\frac12\zeta(1-2r)\right]\\
&\phantom{=}\,+2\tilde{c}\,\sum_{r=2}^{\infty}
\frac{(-1)^{r-1}\binom{2r}{r}h^r x^{2r-1}}{4^r(2r-1)!}
\sum_{j=2}^{r}(-1)^{r-j}\binom{r}{j}\frac{\delta q_{2j-1}}{2j-1}+O(\tilde{c}^{-1}).
\end{aligned}
\]
The part proportional to $M_1-\frac1{24}$ sums to $\frac{M_1-\frac1{24}}{\sqrt h}\sinh(\sqrt h\,x)$. For the part containing $\zeta(1-2r)$, the expansion of cotangent~\cite[Page 12--13, 17]{andrews:1999xpd} gives
\begin{equation}\label{eqt:vacuum-series}
-\frac12\sum_{r=1}^{\infty}
\frac{(-1)^{r-1}\zeta(1-2r)x^{2r-1}}{(2r-1)!}
=-\frac14\cot\frac x2+\frac1{2x},
\end{equation}
where the value at the origin is defined by the limit. To separate the dependence on the given data from that on the parameter $h$, define
\begin{equation}\label{eqt:B}
B(z):=2\tilde{c}\,\sum_{r=2}^{\infty}
\frac{(-1)^{r-1}\binom{2r}{r}z^{2r-1}}{4^r(2r-1)!}
\sum_{j=2}^r(-1)^{r-j}\binom rj\frac{\delta q_{2j-1}}{2j-1}.
\end{equation}
The part containing the charge data is precisely $\sqrt h\,B(\sqrt h\,x)$. Thus,
\begin{equation}\label{eqt:F-from-B}
F(x)=\frac{M_1-\frac{1}{24}}{\sqrt h}\sinh(\sqrt h\,x)+\sqrt h\,B(\sqrt h\,x)-\frac14\cot\frac x2+\frac1{2x}+O(\tilde{c}^{-1}).
\end{equation}
For the finitely supported solution under consideration, moving the $B$ term to one side leaves an expression analytic in $|x|<2\pi$. Hence the series defining $B$ converges in $|z|<2\pi\sqrt h$, and Eq.~(\ref{eqt:F-from-B}) holds in $|x|<2\pi$. Substituting Eqs.~(\ref{eqt:tri}) into Eq.~(\ref{eqt:mf}), we obtain
\begin{equation}\label{eqt:M02-full}
\begin{aligned}
M_{0,\,2}=&\,\frac{2}{\pi}\int_0^\pi\left[\frac{\ell^2q_1-\tilde\Delta}{\sqrt{\frac{\tilde{\Delta}}{\tilde{c}}}}\sinh\!\left(x\sqrt{\frac{\tilde{\Delta}}{\tilde{c}}}\right)-\frac14\cot\frac x2+\frac1{2x}\right.\\
&\,+2\tilde c\,\sum_{r=2}^{\infty}\frac{(-1)^{r-1}}{(2r-1)!}\binom{2r}{r}\left(\frac{\tilde\Delta}{4\tilde c}\right)^r x^{2r-1}\left.\sum_{j=2}^{r}(-1)^{r-j}\binom{r}{j}\frac{\delta q_{2j-1}}{2j-1}\right]^2\,\mathrm{d}x+O(\tilde{c}^{-1}).
\end{aligned}
\end{equation}

Eqs.~(\ref{eqt:reconst}), (\ref{eqt:inversion}) and (\ref{eqt:M02-full}) are well-defined for any large finite large $\ell$. However, this convergence is useless for practical computations. Nevertheless, one can introduce a structure
\begin{equation}\label{eqt:tau}
N_k(\tau)=-2(-1)^k\sum_{r=1}^{\infty}\left[\sum_{m=0}^{r-1}\frac{(-1)^m(\pi k)^{2m}}{(2m+1)!}\right]\frac{M_{2r-1}}{k^{2r-1}}\tau^{2r-1}.
\end{equation}
The right-hand side converges at $\tau\sim\ell^{-3}$ due to $M_{2r-1}\sim\ell^{2r}$, which determines an analytic function $N_k(\tau)$, and then the reconstruction formula is understood as
\begin{equation}
n_k=N_k(1).
\end{equation}
\subsection{Constraint on the highest weight}\label{app:constraint}
In the preceding part of this appendix, we solve a microstate from qKdV charges in each conformal family. However, this reconstruction requires all occupation numbers to be nonnegative, which constrains the highest weight. One knows from Appendix \ref{app:der} that the large central charge $c$ should be treated as a constant. However, a plausible theory should avoid fine-tuning of its fundamental parameters, so we demand robustness against variations in $c$. Based on this natural assumption, we claim the following proposition: The set of all values of qKdV charges statistically determines the highest weight of the conformal family of the microstate, up to thermodynamically subextensive terms. 

For fixed $\tilde{c}$, two parameters $\tilde{\Delta}$ and $h$ form a bijection, so either can be taken as the independent variable. Although the proposition to be established needs to treat the highest weight, or equivalently $\tilde{\Delta}$, as the independent variable and ultimately fixes its value, we recast the argument in terms of $h$ for ease of comparison with earlier formulae in this appendix. 

First of all, the scaling Eq.~(\ref{eqt:Delta}) requires $h>0$. The energy relation determines $M_1=\ell^2q_1-\tilde c h+\frac{1}{24}$, and the triangular inversion formula Eq.~(\ref{eqt:moment-inverse}) then determines all odd moments. By Eq.~(\ref{eqt:F-Taylor}), these moments uniquely determine the odd entire function $F(x)$. Orthogonality of the sine functions further gives Eq.~(\ref{eqt:on}). Thus, for fixed $h$, a finitely supported solution $\{n_k\}$, if it exists, is unique, as we mentioned.

If two solutions have different parameters, denote the larger parameter by $h$ and write $h'=\lambda h$, where $0<\lambda<1$. Denote the corresponding finite sine sums by $F,\,G$, respectively. Since at least one parameter is no smaller than $\frac{1}{24\tilde{c}}$, the larger parameter satisfies $\tilde c h\ge\frac{1}{24}$, and hence
\[
\ell^2q_1=\tilde c h+M_1-\frac1{24}\ge M_1\ge0.
\]

By Eq.~(\ref{eqt:B}), both solutions correspond to the same $B$. Rearranging and rescaling Eq.~(\ref{eqt:F-from-B}) gives
\[
\begin{aligned}
\sqrt{\lambda h}B(\sqrt{\lambda h}\,x)
&=\sqrt\lambda F(\sqrt\lambda x)-\frac{\lambda(\ell^2q_1-\tilde c h)}{\sqrt{\lambda h}}\sinh(\sqrt{\lambda h}\,x)+\frac{\sqrt\lambda}{4}\cot\frac{\sqrt\lambda x}{2}-\frac1{2x}+O(\tilde{c}^{-1})\\
&=G(x)-\frac{\ell^2q_1-\tilde c\lambda h}{\sqrt{\lambda h}}\sinh(\sqrt{\lambda h}\,x)+\frac14\cot\frac x2-\frac1{2x}+O(\tilde{c}^{-1}).
\end{aligned}
\]
Comparing the two expressions, eliminating $B$, we obtain $\exists g(x)=O(\tilde{c}^{-1})$, 
\[
G(x)+g(x)={}\sqrt\lambda F(\sqrt\lambda x)+\frac{\sqrt\lambda}{4}\cot\frac{\sqrt\lambda x}{2}-\frac14\cot\frac x2
+\frac{(1-\lambda)\ell^2q_1}{\sqrt{\lambda h}}\sinh(\sqrt{\lambda h}\,x).
\]
Since Eq.~(\ref{eqt:F-from-B}) holds for $|x|<2\pi$ and $0<\lambda<1$, this identity holds on $[0,\pi]$, with its value at the origin defined by the limit.

Next consider the endpoint $x=\pi$. Using $\left|\sin(k\pi\sqrt\lambda)\right|=\left|\sin(k\pi(\sqrt\lambda-1))\right|\le k\pi(1-\sqrt\lambda)$ and $n_k\ge0$, we obtain
\[
\left|\sqrt\lambda F(\pi\sqrt\lambda)\right|\le\pi\sqrt\lambda(1-\sqrt\lambda)M_1.
\]
Using $\sinh y\ge y$ and $\ell^2q_1\ge M_1\ge0$, we further obtain
\begin{equation}
\begin{aligned}
G(\pi)+g(\pi)&\ge-\pi\sqrt\lambda(1-\sqrt\lambda)M_1+\pi(1-\lambda)\ell^2 q_1+\frac{\sqrt\lambda}{4}\cot\frac{\pi\sqrt\lambda}{2}\\
&\ge\pi(1-\sqrt\lambda)M_1+\frac{\sqrt\lambda}{4}\cot\frac{\pi\sqrt\lambda}{2}>0.
\end{aligned}
\end{equation}
The last step uses $0<\pi\sqrt\frac\lambda2<\frac\pi2$, so the cotangent term is strictly positive.

So, there exists $G(\pi)>0$ at some $\tilde{c}$, contradicting $G(\pi)=0$, which holds because $G$ is a finite sine sum with integer frequencies. Hence $h=h'$, i.e.~$\tilde{\Delta}$ is fixed under the thermodynamic limit. Uniqueness of $\{n_k\}$ for fixed $\tilde{\Delta}$ completes the argument. 

From the scaling Eq.~(\ref{eqt:Delta}), the difference between two highest weights must be at most at the scaling $\ell^2$, where a difference at $\ell^2$ can be definitely characterized by the positive parameter $\lambda\neq1$. Therefore, admissible difference should be zero or subextensive. 

\section{Should $c$ be regarded as a constant or variable in GETH?}\label{app:der}

This appendix clarifies that the central charge should be regarded as a constant instead of a variable in GETH, i.e.~derivative terms of qKdV charges should not be allowed in the expression of quasi-primary expectations for the truncated GETH.

A specific CFT has a fixed central charge $c$, or $\tilde{c}$. This appendix primarily addresses a subtlety pervading the generalized eigenstate thermalization theory, i.e.~whether the $q_{2k-1}$'s can be regarded as functions of $\tilde{c}$, or $\tilde{c}$ must be treated as a given constant instead of a variable.

To touch the essence, we first review the process of deriving the necessity in Section \ref{sec:introduction}.

For reasons of generality, we may obviously first regard $\langle\mathcal{O}\rangle_{\mathrm{GGE}}$ and $\langle\alpha|\mathcal{O}|\alpha\rangle$ as functions of the independent variable $\tilde{c}$. From Eqs.~(\ref{eqt:z}) and (\ref{eqt:ggemu}), we see that the only dependence of ${\rho}_{\mathrm{GGE}}$ on $\tilde{c}$ lies in the eigenstate spectrum $\{|\alpha'\rangle\}$. With Eq.~(\ref{eqt:chempot}), the dependence on $\tilde{c}$ is transferred to $\{q_{2k-1}\}$, i.e.~$\langle\alpha|\mathcal{O}|\alpha\rangle(\tilde{c})$ can be written as a functional of $\{q_{2k-1}(\tilde{c})\}$. More weakly, for any specific $\tilde{c}$, the corresponding $\langle\alpha|\mathcal{O}|\alpha\rangle$ can be written as a functional of $\{q_{2k-1}(\tilde{c})\}$. This is for sure.

Nevertheless, the above derivation makes only weak use of the conditions in the step from $\langle\alpha|\mathcal{O}|\alpha\rangle(\tilde{c})$ to $\langle\alpha|\mathcal{O}|\alpha\rangle$, so we should aim to derive a stronger necessary condition. In this instance, we directly change the initial object of analysis from $\{|\alpha'\rangle(\tilde{c})\}$ to a more strict $\{|\alpha'\rangle\}$, instead of bringing in $\{|\alpha'\rangle\}$ only partway through the analysis. All the dependence of $Z_{\mathrm{GGE}}$ in Eq.~(\ref{eqt:z}) and $\langle\mathcal{O}\rangle_{\mathrm{GGE}}$ in Eq.~(\ref{eqt:ggemu}) on $\tilde{c}$ has been eliminated. Then Eq.~(\ref{eqt:chempot}) can still be solved as before; and the right-hand side of Eq.~(\ref{eqt:gethexp}) is not a function of $\tilde{c}$.

By analyzing the two objects of analysis of differing scopes and comparing the results, we see that the above contribution of the variable $\tilde{c}$ arises purely from parametrizing $\{|\alpha'\rangle\}$ by $\tilde{c}$, i.e.~from artificially extending the scope of consideration to a neighborhood of $\tilde{c}$; $Z_{\mathrm{GGE}}$ and $\langle\mathcal{O}\rangle_{\mathrm{GGE}}$, and then $\langle\alpha|\mathcal{O}|\alpha\rangle$ themselves, should not depend on $\tilde{c}$ at all.

In a nutshell, with $\tilde{c}$ fixed, the thermodynamic theory is already complete: $\rho_{\mathrm{GGE}}$ is already sufficient to determine the full behavior of a thermodynamic system, and introducing any partition function $\rho_{\mathrm{GGE}}(\tilde{c})$ in the neighborhood of $\tilde{c}$ does not add any applicable information, i.e.~no matter how $\rho_{\mathrm{GGE}}(\tilde{c})$ depends on the different neighbor behavior of given $\tilde{c}$, the physics at that $\tilde{c}$ remains unchanged.

Looking at this problem from the sufficiency side is more direct. In Section \ref{sec:introduction} we mentioned a step
\begin{equation}\label{eqt:app}
\langle\mathcal{O}\rangle_{\mathrm{GGE}}=f_{\mathcal{O}}(\{q_{2k-1}^{\mathrm{GGE}}\}).
\end{equation}
This is because $\{q_{2k-1}\}$ takes its value at its peak point $\{q_{2k-1}^{\mathrm{GGE}}\}$. If we consider the dependence of $\{q_{2k-1}\}$ on $\tilde{c}$, i.e.~introduce $\partial_{\tilde{c}}q_{2k-1}$, $\partial_{\tilde{c}}^2q_{2k-1}$, $\cdots$, then Eq.~(\ref{eqt:app}) should be rewritten as
\begin{equation}
\langle\mathcal{O}\rangle_{\mathrm{GGE}}=f_{\mathcal{O}}(\{q_{2k-1}\},\,\{\partial_{\tilde{c}}^nq_{2k-1}\}).
\end{equation}
Under this circumstance, $\{q_{2k-1}\}$ can still take its value $\{q_{2k-1}^{\mathrm{GGE}}\}$, but $\{\partial_{\tilde{c}}^2q_{2k-1}\}$ has no corresponding equations; instead, it is uniquely determined by $\{q_{2k-1}\}$, and in general it may not attain the statistical peak point. Therefore, the whole formulation fails.

In summary, the unmodified GETH based on the GGE Eq.~(\ref{eqt:rhogge}) only allows the expectations of local observables of the vacuum family to be expressed in terms of the exact qKdV charges themselves, but not merely in terms of qKdV charges treated as functions of $\tilde c$.

\bibliographystyle{JHEP}
\bibliography{reference}

\end{document}